\global\def\draftcontrol{0}

   \def\versionno{Phase transition }
\catcode`\@=11

\expandafter\ifx\csname draftcontrol\endcsname\relax\global\def\draftcontrol{0}
\fi

{\count255=\time\divide\count255 by 60
\xdef\hourmin{\number\count255}
\multiply\count255 by-60\advance\count255 by\time
\xdef\hourmin{\hourmin:\ifnum\count255<10 0\fi\the\count255}}
\def\draftdate{\number\month/\number\day/\number\year\ \ \ \hourmin }

\newcommand\makepapertitle{\par
  \begingroup
    \renewcommand\thefootnote{\@fnsymbol\c@footnote}%
    \def\@makefnmark{\rlap{\@textsuperscript{\normalfont\@thefnmark}}}%
    \long\def\@makefntext##1{\parindent 1em\noindent
            \hb@xt@1.8em{%
                \hss\@textsuperscript{\normalfont\@thefnmark}}##1}%
     \newpage
     \global\@topnum\z@   % Prevents figures from going at top of page.
     \@makepapertitle
     \thispagestyle{empty}\@thanks
  \endgroup
  \setcounter{footnote}{0}%
  \global\let\thanks\relax
  \global\let\makepapertitle\relax
  \global\let\@makepapertitle\relax
  \global\let\@thanks\@empty
  \global\let\@author\@empty
  \global\let\@date\@empty
  \global\let\@title\@empty
  \global\let\title\relax
  \global\let\author\relax
  \global\let\date\relax
  \global\let\and\relax
  \def\version{\let\version\@version\@gobble}
}
\def\@makepapertitle{%
  \newpage
   \ifnum\draftcontrol=1 {}
   \version\versionno
   \vskip 3em%
   \else
   \hfill\hbox to 3cm {\parbox{4cm}{\@pubnum}\hss}%
   \vskip 3em%
   \fi
   \begin{center}%
   \let \footnote \thanks
     {\LARGE {\@title}}%
     \vskip 1.5em%
     {\normalsize%\large
       \lineskip .5em%
       \begin{tabular}[t]{c}%
         \@author
       \end{tabular}\par}%
     \vskip 1.5em%
     {\@bstract}%
     \end{center}%
     \vskip 1.5em
     \@date%
   \par
}

\gdef\@pubnum{}
\def\pubnum#1{%
  \gdef\@pubnum{#1}}

\gdef\@bstract{}
\def\Abstract#1{%
  \gdef\@bstract{%
   \parbox{\textwidth-0pc}{%
   \centerline{\bf Abstract}\penalty1000%
\noindent%\abstractfont \baselineskip=12pt
\renewcommand\baselinestretch{1.0}%
{#1}}}
}

\def\ps@paper{\let\@mkboth\@gobbletwo%
     \ifnum\draftcontrol=1
        \def\@oddfoot{\hbox to \textwidth{\tiny \versionno \hfil\tiny\draftdate}%
        \hskip -\textwidth \hbox to \textwidth{\hfil\rm\thepage\hfil}}%
     \else\def\@oddfoot{\hbox to \textwidth{\hfil\rm\thepage\hfil}}
     \fi
     \let\@evenfoot\@oddfoot
}
\def\@version#1{\ifnum\draftcontrol=1
\typeout{}\typeout{#1}\typeout{}
\vskip3mm\centerline{\hbox{\fbox{\normalsize{\tt DRAFT -- #1 -- }
                   {\draftdate}}}}\vskip3mm
\fi}
\let\version\@version
\long\def\eqlabel#1{\ifnum\draftcontrol=1
                    \tag@false  % there are some problems with multline without this
                    \tag*{(\theequation) \hbox to -0.2cm{\hspace{0cm}\small{#1}\hss}}
                    \refstepcounter{equation}
                    \edef\@currentlabel{\theequation}
                    \ltx@label{#1}          % use old LaTeX \label instead of new definition
                    \else
                    \label{#1}
                    \fi
                    }
\let\st@bibitem\@bibitem
\let\st@lbibitem\@lbibitem
\ifnum\draftcontrol=1
  \def\@bibitem#1{%
    \st@bibitem{#1}\a@@label{#1}\ignorespaces}
  \def\@lbibitem[#1]#2{%
    \st@lbibitem[#1]{#2}\a@@label{#2}\ignorespaces}
  \def\a@@label#1{%
    \gdef\a@lab{\smash{\normalfont\small#1}}
    \ifvmode
      \if@inlabel
        \global\setbox\@labels\hbox{%
          \llap{\a@lab\let\a@lab\relax
                \kern\@totalleftmargin\kern\marginparsep}%
          \box\@labels}%
      \fi
    \fi}
\fi
\documentclass[12pt,letterpaper]{article}
\usepackage{amsmath,amssymb,array,calc,rotating,epsfig,psfrag,graphicx,subfigure}
\usepackage[nosort]{cite}
\ifnum\draftcontrol=1
\fi
\renewcommand\baselinestretch{1.25}
\renewcommand\section{\@startsection {section}{1}{\z@}%
                                   {-3.5ex \@plus -1ex \@minus -.2ex}%
                                   {2.3ex \@plus.2ex}%
                                   {\normalfont\large\bfseries}}
\renewcommand\subsection{\@startsection{subsection}{2}{\z@}%
                                   {-3.25ex\@plus -1ex \@minus -.2ex}%
                                   {1.5ex \@plus .2ex}%
                                   {\normalfont\normalsize\bfseries}}
\renewcommand\subsubsection{\@startsection{subsubsection}{3}{\z@}%
                                   {-3.25ex\@plus -1ex \@minus -.2ex}%
                                   {1.5ex \@plus .2ex}%
                                   {\normalfont\normalsize\it}}
\renewcommand\paragraph{\@startsection{paragraph}{4}{\z@}%
                                   {-3.25ex\@plus -1ex \@minus -.2ex}%
                                   {1.5ex \@plus .2ex}%
                                   {\normalfont\normalsize\bf}}

\def\revise#1       {\raisebox{-0em}{\rule{3pt}{1em}}%
                     \marginpar{\raisebox{.5em}{\vrule width3pt\
                     \vrule width0pt height 0pt depth0.5em
                     \hbox to 0cm{\hspace{0cm}{%
                     \parbox[t]{4em}{\raggedright\footnotesize{#1}}}\hss}}}}

\def\sqr#1#2{{\vcenter{\vbox{\hrule height.#2pt
 \hbox{\vrule width.#2pt height#1pt \kern#1pt
 \vrule width.#2pt}\hrule height.#2pt}}}}

\catcode`\@=12

\usepackage{color}

\newcommand{\be}{\begin{equation}}
\newcommand{\ee}{\end{equation}}
\newcommand{\beq}{\begin{equation}}
\newcommand{\eeq}{\end{equation}}
\newcommand{\ba}{\begin{eqnarray}}
\newcommand{\ea}{\end{eqnarray}}

\def\lbldef#1#2{\expandafter\gdef\csname #1\endcsname {#2}}

\def\eqalign#1{\vcenter{\openup1\jot   }}

\def\href#1#2{#2}

\newcommand{\ber}{\begin{eqnarray}}
\newcommand{\eer}{\end{eqnarray}}

\newcommand{\bea}{\begin{eqnarray}}
\newcommand{\eea}{\end{eqnarray}}

\newcommand{\beqar}{\begin{eqnarray}}

\newcommand{\eeqar}{\end{eqnarray}}
\newcommand{\dsl}
  {\kern.06em\hbox{\raise.15ex\hbox{$/$}\kern-.56em\hbox{$\partial$}}}

\newcommand{\eeqarr}{\end{eqnarray}}
\newcommand{\ZZ}{{\rm \kern 0.275em Z \kern -0.92em Z}\;}

\makeatletter \@addtoreset{equation}{section} \makeatother
\renewcommand{\theequation}{\thesection.\arabic{equation}}
\def\be{\begin{equation}}
\def\ee{\end{equation}}
\def\bea{\begin{eqnarray}}
\def\eea{\end{eqnarray}}

\def\ba{\bar{\alpha}}

\def\bear{\begin{eqnarray}}
\def\eear{\end{eqnarray}}

\newcommand\belabel[1]{\begin{equation}\label{#1}}

\newcommand\bearlabel[1]{\begin{eqnarray}\label{#1}}

\begin{document}

\begin{titlepage}

\version\versionno

\vskip -.8cm

\rightline{\small{\tt MCTP-07-30}}

\vskip 1.7 cm

\centerline{\bf \Large Holographic Entanglement Entropy at Finite Temperature  }

\vskip .4cm
%\centerline{\bf \Large and Phase Transitions }

\vskip .7 cm

%\centerline{\bf \Large }

\vskip .2cm \vskip 1cm

\centerline{\large Ibrahima Bah$^1$, Alberto Faraggi$^{1}$,}

\vskip .5cm

 \centerline{\large Leopoldo A. Pando Zayas$^{1}$ and C\'esar A. Terrero-Escalante$^{2}$ }

\vskip 1cm

\vskip .5cm
\centerline{\it ${}^1$ Michigan Center for Theoretical
Physics}
\centerline{ \it Randall Laboratory of Physics, The University of
Michigan}
\centerline{\it Ann Arbor, MI 48109-1040}

\vspace{1cm}
\centerline{\it ${}^2$ Departamento de F\'{\i}sica, Centro de Investigaci\'on y Estudios Avanzados del IPN,}
\centerline{ \it Apdo. Postal 14-740, 07000 M\'exico D.F., M\'exico}

\vspace{1cm}

\begin{abstract}
Using a holographic proposal for the entanglement entropy we study its behavior in various supergravity backgrounds. We are particularly interested in the possibility of using the entanglement entropy as way to detect transitions induced by the presence horizons. We consider several geometries with horizons: the black hole in $AdS_3$, nonextremal Dp-branes, dyonic black holes asymptotically to $AdS_4$ and also Schwarzschild black holes in global $AdS_p$ coordinates. Generically,  we find that the entanglement entropy does not exhibit a transition, that is, one of the two possible configurations always dominates. 
%Our results suggest that the entanglement entropy  is not a good  order parameter for confinement/deconfinement transition.
\end{abstract}

%\end{center}

%\noindent

\end{titlepage}

%\newpage

%--------+---------+---------+---------+---------+---------+---------+
%Body

%%%%%%%%%%%%%%%%%%%%%%%%%%%%%%%%%%%%%%%%%%%%%%%%%%%%%%%%%%%%%%%%%%%%%%%%%%%
\section{Introduction}
%%%%%%%%%%%%%%%%%%%%%%%%%%%%%%%%%%%%%%%%%%%%%%%%%%%%%%%%%%%%%%%%%%%%%%%%%%
Given a system in a pure quantum state $|\Psi>$ and density matrix $\rho= |\Psi><\Psi|$, if
we split the system into two subsystems $A$ and $B$,
the reduced density matrix is obtained by
tracing over the degrees of freedom in the complementary subsystem, say, $\rho_A= {\rm Tr}_B \rho$.
The entanglement entropy is defined as the von Neumann entropy
\be
S_A=-{\rm Tr}\rho_A\log \rho_A.
\ee
This provides a measure of how entangled or ``quantum''  a system is. Entanglement plays a central role
in quantum information theory as it determines the ability to send
quantum information  \cite{book1}. The entanglement entropy also plays an important role in the study of
strongly correlated quantum systems \cite{osbornenielsen}.

The above definition is completely field theoretical. Interestingly in the context where such field theories have supergravity duals, a prescription for the holographic computation of the entanglement entropy has been provided \cite{Ryu:2006bv},that is, a prescription for computing $S_A$ completely within the holographic gravity dual.
Inspired by the Bekenstein-Hawking entropy, it suggests to calculate the entanglement entropy as the
area  associated to a minimal surface $\gamma_A$  whose boundary is the region $A$ in the field theory living at the
boundary:
\be
S_A=\frac{{\rm Area}\, {\rm of}\, \gamma_A}{4G_N^{d+2}}.
\ee
This recipe has been successfully applied to various systems and extended in different directions \cite{Ryu:2006ef} including
its covariant formulation \cite{Hubeny:2007xt}.
A slightly modified version of the entanglement entropy is
\be
S_A=\frac{1}{4G_N^{(10)}}\int d^8\sigma e^{-2\phi}\sqrt{G_{ind}^{(8)}}.
\ee
The entropy is obtained by minimizing the above action over all surfaces that approach the boundary of the subsystem $A$.
In a very interesting paper \cite{Klebanov:2007ws}, it was suggested that in the presence of regions with
collapsing cycles, which are
typical for supergravity duals of confining field theories, alternative surfaces arise 
(see also \cite{Nishioka:2006gr}). 
By comparing the entropy due to two
different configurations, it was shown that the entanglement entropy  could be an
order parameter for the confinement/deconfinement transition. The motivation comes from the fact that
the entanglement entropy jumps from a
configuration with dominant term of the form $N^2$ to a configuration with leading term $N^0$ for supergravity backgrounds describing theories with $SU(N)$ gauge
groups in the large $N$ and fixed 't Hooft coupling limit.

One intriguing fact about the results of \cite{Klebanov:2007ws} is that, by analyzing a surface in the supergravity dual
to the confined phase of the field theory, one is able to anticipate the existence of a deconfined phase. This
prompts the natural question
of whether the exploration of the deconfined phase can equally well give information about the existence of a
confined phase.

According to the AdS/CFT correspondence \cite{Maldacena:1997re}, the dual of field theories at finite temperature involves
black hole horizons on the supergravity side \cite{Witten:1998zw}.
The horizon provides a natural end of the space similar to the situation discussed in \cite{Klebanov:2007ws}.
In fact, in the context of the AdS/CFT is has already been established in
various situations that there are phase transitions associated to different behavior of surfaces describing branes in the presence
of a horizon \cite{Aharony:2006da,Parnachev:2006dn,Babington:2003vm,Mateos:2006nu,Albash:2006ew,Kobayashi:2006sb}.

\begin{figure}[!h]
\centering
%\hfill
\includegraphics[width=0.45\linewidth]{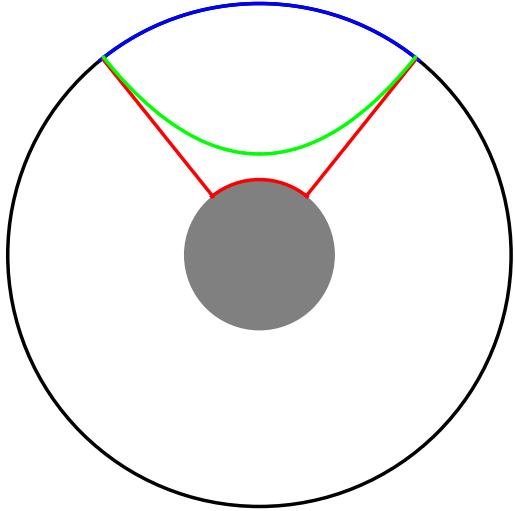}\\
\caption{Two competing configurations for the entanglement entropy in the presence of a black hole horizon. The green
surface represents a continuous configuration while the red surface goes straight down from the
boundary to the horizon.  The subsystem $A$ is given by the blue section.  Its characteristic length is $l$.}\label{fig:discon}
\end{figure}

In this paper we explore the entanglement entropy for various field theories at finite temperature using its holographic definition. Namely, we study the entanglement entropy for various black hole geometries. We define our subsystem $A$ to be roughly determined by an interval of length $l$ on a curve generated by a spacelike killing vector in the conformal boundary of the geometry. Generically, there are two surfaces that satisfy the boundary conditions: a smooth one and a piece-wise smooth (see figure \ref{fig:discon}). We study the behavior of these two surfaces as a function of the distance $l$.

In section \ref{2d} we discuss the gravity dual of 2D CFT at finite temperature -- the BTZ black hole --  where the smooth surface is just the geodesic anchored on $A$.  The piece-wise smooth surface is the curve composed of the lines stretching from the conformal boundary to the horizon connected by the segment of length $l$ on the horizon. We show that the smooth configuration is always dominant from the point of view of the thermodynamical comparison. Section \ref{general} presents a general setup for the computations at hand and discusses explicitly the case of the the supergravity  backgrounds describing the ${\cal N}=4$ plasma and other field theories at finite temperature.
Also in section \ref{general} we extend the analysis to black $p$-branes corresponding to various gauge theories on the world volume of $p$-branes at
finite temperature. Using the supergravity backgrounds for nonextremal Dp-branes for all values of $p$, we found that the entanglement entropy is given by the smooth surface for all $l$, except for $p=6$.  In the $p=6$ case we observe that the entanglement entropy is given by the smooth surface at large $l$ and by the piece-wise smooth surface at small $l$.  Furthermore in these geometries, we observe that the radius of the black hole factors as an overall scaling factor.  This scaling implies that the entanglement entropy has the same $N$ (number of color) dependence for all the temperatures. In section \ref{globalads}, we study the entanglement entropy in black hole geometries in global $AdS_5$ and $AdS_4$.  The main motivation for our analysis comes from the fact that the Hawking-Page phase transition takes place only in  global coordinates \cite{Hawking:1982dh}, that is, for the nonextremal Dp-brane geometries there is no Hawking-Page phase transition. In the context of the AdS/CFT this is interpreted as a transition for the dual field theory on a sphere  \cite{Witten:1998zw}. Since this feature is absent in the Poincare patch we are motivated to study the entanglement entropy in global AdS backgrounds. However, we find that in global coordinates,  the entanglement entropy is given also by the smooth surface.

Our work suggests 
that starting from the deconfined phase, changing the lenght of the subsystem does not allow for a transition into confinement, as opposed to what was observed in \cite{Klebanov:2007ws,Nishioka:2006gr}.
According with their results, at zero temperature the length of the subsystem seems to play the role of temperature in the confined phase, while in our case this lenght corresponds possibly to some other thermodynamical quantity.
%that the entanglement entropy is not a good order parameter for confinement/deconfinement as suggested by \cite{Klebanov:2007ws}. 
It is probably worth mentioning  a recent study of the topological entanglement entropy \cite{Pakman:2008ui} where no change in the dependence on $N$ was observed. However there is mounting evidence that the entanglement entropy is a {\it bona fide} thermodynamical quantity and its precise meaning should be illuminated through further work.  In section \ref{conclusions} we discuss some of the directions suggested by our exploration of the entanglement entropy in the context of finite temperature.

%%%%%%%%%%%%%%%%%%%%%%%%%%%%%%%%%%%%%%%%%%%%%%%%%%%%%%%%%%%%%%%%%%%%%%%%%%%
\section{2D CFT at finite temperature from the BTZ black hole }\label{2d}
%%%%%%%%%%%%%%%%%%%%%%%%%%%%%%%%%%%%%%%%%%%%%%%%%%%%%%%%%%%%%%%%%%%%%%%%%%%
We begin this section with a discussion of the holographic entanglement entropy using the gravitational background dual to a 2D CFT at finite temperature. Most of the calculation was explicitly done in \cite{Ryu:2006ef}, however, we reproduce it here making emphasis on the thermodynamical competition between the two configurations, something that was not considered in \cite{Ryu:2006ef}.
The relevant geometry holographically describing the field theory at finite temperature is the BTZ black hole
\cite{Banados:1992wn}:

\be
ds^2=-(r^2-r_+^2)dt^2 + \frac{R^2}{r^2-r_+^2}dr^2 +r^2 d\phi^2.
\ee
The smooth surface is parametrized by $t=$constant and $r=r(\phi)$.  Subsystem A is defined as the region given by $0\leq \phi \leq 2 \pi l/L$ where $L$ is the circumference of the boundary.  Thus, $l$ is the characteristic size of A.  For the BTZ black hole, the smooth surface is just the geodesic in the bulk that connects the two boundary points of A.  The action of this curve is given by
\be
A_c=\int d\phi \sqrt{r^2+\frac{R^2}{r^2-r_+^2}r'^2}.
\ee
The equation of motion can be integrated to give
\be
\frac{dr}{d\phi}=\frac{r}{R r_*}\sqrt{(r^2-r_*^2)(r^2-r_+^2)}.
\ee
This allows us to relate the length in the $\phi$ direction with the minimum of the curve, $r_*$:
\be
\frac{2\pi l}{L}=\frac{R}{r_+}\ln \frac{r_*+r_+}{r_*-r_+}.
\ee
The temperature is $\beta=RL/r_+$. The gravitational theory on $AdS_3$ with radius $R$
is dual to a 2D CFT living on its boundary with central charge $c=3R/2G_N^{(3)}$  \cite{Brown:1986nw}.

The length of the geodesic is then
\be
Area_c=2R\ln \frac{r_\infty}{r_+}\sinh\left(\frac{r_+}{RL}\pi l \right)
\ee
where $r_\infty$ is a UV cutoff.

The entanglement entropy given by the continuous configuration is:
\be
S_{c}=\frac{1}{4G_N^{(3)}}Area_{c}=\frac{2R}{4G_N^{(3)}}\ln \frac{r_\infty}{r_+}\sinh\left(\frac{r_+}{RL}\pi l \right).
\ee
As noted in \cite{Ryu:2006bv}, this result is in agreement with the field theory result for a
2D CFT at finite temperature \cite{Calabrese:2004eu}:
\be
S_A=\frac{c}{3}\log\left(\frac{\beta}{\pi a}\sinh \frac{\pi l}{\beta}\right),
\ee
where $a$ is a ultraviolet cutoff that can be thought of as a lattice spacing.

The piece-wise smooth surface is given by the parametrization gluing the 3 surfaces  described as by $\phi = 0$, $r=r_+$ and $\phi = 2 \pi l/L$.  The area for this curve is then
\be
Area_d=2 \int\limits_{r_+}^{r_\infty} \frac{R}{\sqrt{r^2-r_+^2}}dr + \frac{2r_+  \pi l}{L}
=2 R \ln \frac{r_\infty}{r_+} + \frac{2r_+  \pi l}{L}.
\ee
The first term of the piece-wise smooth configuration consists of two lines that go from the boundary to the horizon, their contribution to the entropy is:
\be
S_{d(1)}=\frac{1}{4G_N^{(3)}}2\int\limits_{r_+}^{r_\infty} \frac{R}{\sqrt{r^2-r_+^2}}dr
= \frac{R}{2G_N^{(3)}}\ln \frac{r_\infty}{r_+}.
\ee
By construction, this entropy is independent of the length $l$ of the interval that defines the subsystem $A$. In intrinsic
field theoretic terms we could rewrite it as:
\be
\eqlabel{discontinuous}
S_{d(1)}=\frac{c}{3}\ln \frac{\beta}{a}.
\ee
The study of entanglement entropies in 1+1 system is well developed. In \cite{Holzhey:1994we} a formula for the
entanglement entropy of a system of length $l$ was obtained for conformal field theories: $(c/3)\ln(l/a)$. Some extensions of
this result have been discussed recently, including systems  away from criticality
\cite{Calabrese:2004eu,Vidal:2002rm,Calabrese:2005zw}.
Systems close to a phase transition (large but finite correlation length) are described
by massive quantum field theories with mass inversely proportional to the correlation length \cite{Calabrese:2004eu}:
\be
S_A\sim \frac{c}{3}\ln\left(\frac{\xi}{a}\right).
\ee
The correlation length that appears above is considered to be the inverse of the mass $m=\xi^{-1}$.
The correlation length $\xi$ and the inverse temperature $\beta$ can be naturally identified in our setup
completing the matching of the gravity (\ref{discontinuous}) with the field theoretic one. Interpreting the second term in the piece-wise smooth configuration is more challenging, we simply note that in field theoretic terms the contribution coming directly from the horizon takes the form: $S_{d(2)}=(c/3) (\pi l/\beta)$.

Now consider the difference in area:
\be
\Delta A = A_c - A_d = 2 R \ln \left[\sinh\left(\frac{r_+  \pi l}{R L}\right)\right] - \frac{2r_+  \pi l}{L}.
\ee
As expected, the difference is UV finite.  Setting it to zero we obtain:
\be
\sinh(x) = e^x \;\;\mbox{where} \;\; x =\frac{r_+  \pi l}{R L}.
\ee
This equation has no solutions and it follows that
\be
A_c < A_d .
\ee
This implies that the entropy is always given by the smooth surface. As noted in \cite{Ryu:2006ef} the answer matches precisely the field theory calculation \cite{Calabrese:2005zw}. More impressive agreement has also been found in the context of disconnected segments in the field theory \cite{Hubeny:2007re}.

%%%%%%%%%%%%%%%%%%%%%%%%%%%%%%%%%%%%%%%%%%%%%%%%%%%%%%%%%%%%%%%%%%%%%%%%%%%
\section{Entanglement entropy for nonextremal Dp branes}\label{general}
%%%%%%%%%%%%%%%%%%%%%%%%%%%%%%%%%%%%%%%%%%%%%%%%%%%%%%%%%%%%%%%%%%%%%%%%%%%

The next geometries that we study are static geometries with a spacelike killing vector $X^\mu$ that commutes with all other killing vectors.  Furthermore we assume that these geometries have conformal radius $r$.  Thus there exist a frame where the metric is given as,
\begin{equation}
ds^2 = -\tau(r) dt^2 + f(r) dr^2 + g(r) dx^2 + h_{ij}dy^i dy^j
\end{equation} where $x$ is the affine parameter along $X^\mu$ and $y^i$ are coordinates of the internal submanifold.  In general, this submanifold can have compact and non-compact directions and $h_{ij}$ can have non-trivial dependence on the coordinates.  However, the geometries considered here satisfy:
\begin{equation}
h_{ij}dy^i dy^j = h^1_{ab}(r)du^a du^b + h^2_{mn}(r,\theta) d\theta^m d\theta^n
\end{equation} where the $u^i$'s are non-compact coordinates and $\theta^m$'s are compact.
The conformal boundary is obtained by taking the large $r$ limit.  The region $A$ that we will consider is parametrized by:
\begin{equation}
t = \mbox{constant} \;\;\; \mbox{and} \;\;\; -\frac{l}{2} < x < \frac {l}{2} \nonumber
\end{equation} where $l$ is the size of the subsystem  $A$.
As discussed above, there are two surfaces that satisfy the minimal surface condition.  The smooth surface is given by,
\begin{equation}
t = \mbox{constant}, \;\;\; \mbox{and} \;\;\; r = r(x) \;\;\; \mbox{with b.c.} \;\; r(\pm l/2) = r_\infty,
\end{equation}  where $r_\infty$ is the location of the holographic boundary.  The second surface is piece-wise smooth and parametrized as:
\begin{equation}
t= \mbox{constant}, \;\;\; \mbox{and} \begin{cases} &x= -\frac{l}{2}  \\ &r= r_0 \\ &x= \frac{l}{2} \end{cases}
\end{equation} where $r_0$ is the lower bound of $r$.  Now we can proceed to study the difference in area of these surfaces.

%%%%%%%%%%%%%%%%%%%%%%%%%%%%%%%%%%%%%%%%%%%%%%%%%
\subsection{Two branches of the holographic entropy}

We start by computing the area of the smooth surface.  The induced metric is:
\begin{equation}
ds^2 =  g(r)(1+\frac{f(r)}{g(r)} r'^2(x)) dx^2 + h_{ij}dy^i dy^j
\end{equation} where prime denotes derivative with respect to $x$.  The volume element is then
\begin{equation}
\sqrt{G_{ind}} = \sqrt{g(r)(1 + \frac{f(r)}{g(r)} r'^2(x)) h^1(r) h^2(r,\theta)}
\end{equation} where $h^1$ and $h^2$ are the determinant of the metric for the non-compact and compact sub-manifolds, respectively.  Since the non-compact sub manifold will contribute an infinite volume, we must work with the area density.  It is given by
\begin{equation}
A = \int dx e^{-2\phi} \sqrt{g(r)(1 + \frac{f(r)}{g(r)} r'^2(x)) h^1(r)} \int d^n \theta \sqrt{h^2(r,\theta)}
\end{equation} where $\phi$ is the dilaton.  Before we proceed, we define the following quantities:
\begin{equation} \sqrt{H(r)} = e^{-2\phi} \sqrt{g(r) h^1(r)} \int d^n \theta \sqrt{h^2(r,\theta)} \;\;\; \mbox{and} \;\;\; \beta^2 = \frac{f(r)}{g(r)}.
\end{equation} The area is:
\begin{equation}
A = \int dx \sqrt{H(r)} \sqrt{1 + \beta^2 r'^2(x)}.
\end{equation} Now we want to find the $r(x)$ that minimizes the area.  Since the Lagrangian does not explicitly depend on $x$, we can use the conserve Hamiltonian to integrate the equation of motion. It is then:
\begin{equation} E = L - \frac{\partial L}{\partial ( r')}( r') = \frac{\sqrt{H(r)}}{\sqrt{1+ \beta^2 r'^2}}.
\end{equation} The constant $E$ can be obtained by considering the turning point $r_*$ where $r'=0$.  We thus obtain
\begin{equation}
E = \sqrt{H(r_*)}.
\end{equation} Physically, $r_*$ is the minimum of the surface.  It corresponds to $x=0$ since the surface must be symmetric under $x \to -x$.  The quantity, $r_*$, can be used to label different surfaces for different values of $l$. This relationship is obtained by integrating the equation of motion one more time,
\begin{equation}
l(r_*) = \int dx = \int \frac{dr}{r'} = 2 \sqrt{H(r_*)} \int_{r_*}^{r_\infty} \frac{\beta(r)}{\sqrt{H(r) - H(r_*)}} dr . \label{l1:1}
\end{equation}
Similarly, we can obtain the area as an integral over $r$,
\begin{eqnarray}
A_c (r_*) &=& \int dx \sqrt{H(r)} \sqrt{1 + \beta^2 r'^2(x)} = \int \frac{dr}{r'} H(r) \nonumber \\
&=& 2 \int_{r_*}^{r_\infty} \frac{\beta(r) H(r)}{\sqrt{H(r) - H(r_*)}}.
\end{eqnarray}
We also compute the area of the piece-wise smooth surface.  The induced line elements for different segments are
\begin{eqnarray}
ds^2 &=& f(r) dr^2  + h_{ij}dy^i dy^j  \;\;\; \mbox{for} \;\;\; x = \pm \frac{l}{2}\\
ds^2 &=&  g(r_0) dx^2 + h_{ij}dy^i dy^j  \;\;\; \mbox{for} \;\;\; r = r_0.
\end{eqnarray} The area of the piece-wise smooth is then
\begin{equation}
A_d = 2 \int_{r_0}^{r_\infty} \beta(r) \sqrt{H(r)} dr + l \sqrt{H(r_0)}.
\end{equation}
In what follows it suffices to identify $H$ and $\beta$ from the geometries; the difference in area and $l$ are given by;

\begin{eqnarray}
 \Delta A (r_*) &=& 2 \int_{r_*}^{r_\infty} \frac{\beta(r) H(r)}{\sqrt{H(r) - H(r_*)}} \left[1 -\left(\frac{H(r_0)}{H(r_*)}\right)^{1/2} \frac{H(r_*)}{H(r)}- \left(1 -  \frac{H(r_*)}{H(r)}\right)^{1/2}\right] \nonumber \\  &-& 2\int_{r_0}^{r_*} \beta(r)\sqrt{H(r)} \\
l(r_*) &=& 2 \sqrt{H(r_*)} \int_{r_*}^{r_\infty} \frac{\beta(r)}{\sqrt{H(r) - H(r_*)}} dr .
\end{eqnarray} It should be noted here that when $r_* = r_0$, the difference in area is negative since
\begin{eqnarray}
\Delta A(r_0) &=& 2 \int_{r_*}^{r_\infty} \frac{\beta(r) H(r)}{\sqrt{H(r) - H(r_*)}} \left[1 - \frac{H(r_0)}{H(r)}- \left(1 -  \frac{H(r_0)}{H(r)}\right)^{1/2}\right]<0.
\end{eqnarray}
Similarly if $\beta \to 0$ fast enough as $r \to r_\infty$, then
\begin{equation}
\Delta A(r_* \to r_\infty) \approx - 2\int_{r_0}^{r_\infty} \beta(r)\sqrt{H(r)} <0.
\end{equation} This statement is almost always true.  We will see an exception where $\beta \propto \frac{1}{\sqrt{r}}$.  Thus we obtain two important results before we even start to look at specific geometries
\begin{equation}
 \Delta A(r_*=r_0) < 0, \;\;\; \mbox{and} \;\;\; \Delta A(r_* \rightarrow r_\infty) < 0. \label{alim}
\end{equation} Thus, if the difference in area is monotonous, the smooth surface gives the entanglement entropy.  Now we proceed to study geometries corresponding to holographic duals of the ${\cal N}=4$ plasma at finite temperature and other field theories leaving in Dp-branes world volumes
and dyonic black hole.

%%%%%%%%%%%%%%%%%%%%%%%%%%%%%%%%%%%%%%%%%%%%%%%%%%%%%%%%%%%%%%%%%%%%%%%%%%%
\subsection{Entanglement entropy in the ${\cal N}=4$ plasma}
%%%%%%%%%%%%%%%%%%%%%%%%%%%%%%%%%%%%%%%%%%%%%%%%%%%%%%%%%%%%%%%%%%%%%%%%%%%
The holographic background dual to the ${\cal N}=4$ plasma is a stack of nonextremal D3 branes. This background has proved to be an interesting playground for finite temperature field theories, in particular, it seems to catch some features displayed by experiments at the Relativistic Heavy Ion Collider. The corresponding background metric is:
\be
ds^2 =R^2\bigg[\frac{du^2}{hu^2}+ u^2 \left(-hdt^2 +dx_idx^i\right) + d\Omega_5^2\bigg],
\ee
with
\be
h=1-\frac{u_0^4}{u^4},
\ee
where the parameter $u_0$ completely characterizes the temperature of the background: $u_0=\pi R\,T$.  The coordinate $u$ is related to the holographic coordinate $r$ by $r=R u$ where $R$ is the AdS radius.
The quantities $H$ and $\beta$ are:
\be
H(u) = (R^8 \Omega_5)^2 u^6, \;\;\; \beta^2(u) = \frac{1}{u^4 - u_0^4}.
\ee
The difference in area and $l$ are:
\begin{eqnarray}
\Delta A &=& 2I u_0^2 \left[\int_{y_*}^{y_{\infty}}\frac{y^{6} -y^3 \sqrt{y^{6}-y_*^{6}} - y_*^3}{\displaystyle{\sqrt{\left(y^{6}-y_*^{6}\right)\left(y^4-1\right)}}}dy -\frac{\sqrt{y_*^4-1}}{2} \right] \\
l &=& \frac{2}{u_0} \int_{y_*}^{y_\infty} \frac{y_*^3 dy}{\sqrt{(y^4-1)(y^6-y_*^6)}} \label{eq:n4l}
\end{eqnarray} where $y=u/u_0$ and $I=\Omega_5 R^8$.  The $y$ coordinate makes it clear that $\Delta A$ is just scales with respect to the temperature. We numerically plot $\Delta A$ as function of $lu_0$ in units of $2Iu_0^2$.

\begin{figure}[!h]
		\subfigure[]{\label{fig:1-a}\includegraphics[scale=.70]{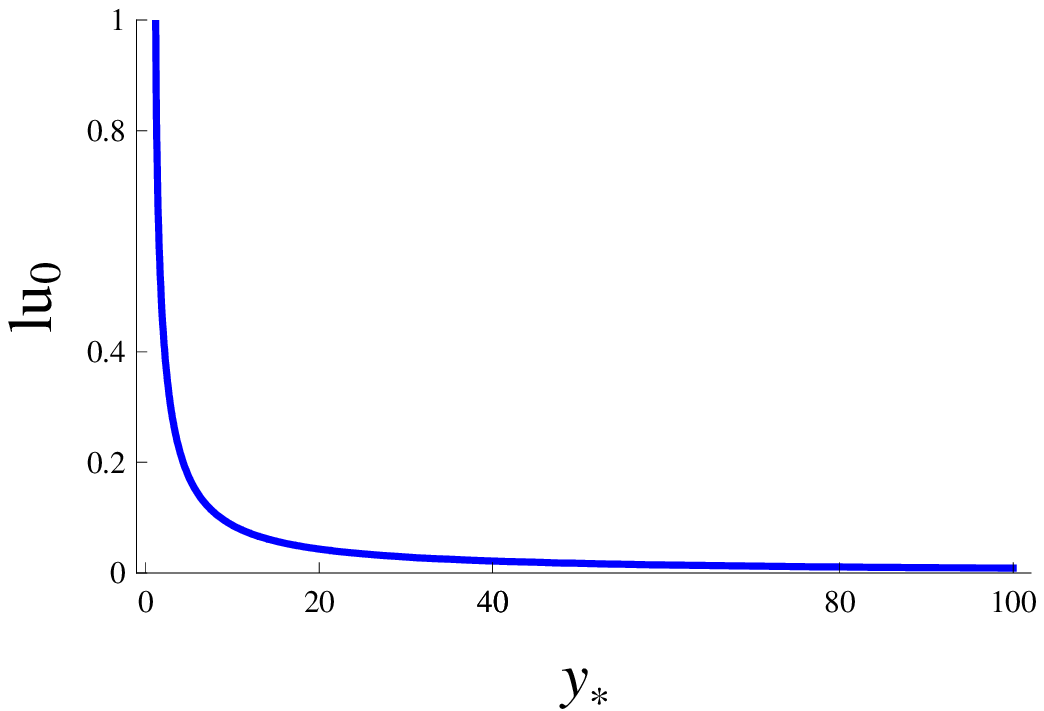}}
		\subfigure[]{\label{fig:1-b}\includegraphics[scale=.70]{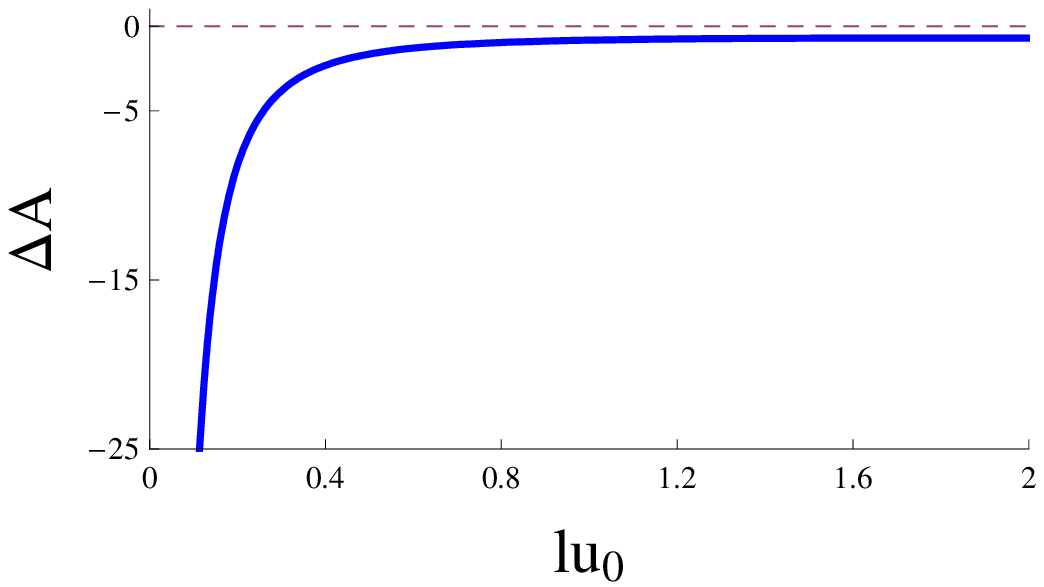}}
	\caption{Figure \ref{fig:1-a}  shows the behavior of  $l$ as a function of $y_*$ for constant $u_0$.  Figure \ref{fig:1-b} Shows the difference in area as a function of $l$.  These plots show no phase transition the entropy in the nonextremal D3 brane background.}
	\label{fig:n4}	
\end{figure}

From figure \ref{fig:n4} we see that the difference in entropy does not change sign; instead it approach 0.  $\Delta A$ is also monotonous  with respect to $l$ as observed from \ref{fig:n4}.  $l$ is monotonous with respect $y_*$; in fact it logarithmically diverges as $y_* \to 1$ as observed from \ref{eq:n4l}.  Thus by \ref{alim}, $\Delta A$ is always negative.

%%%%%%%%%%%%%%%%%%%%%%%%%%%%%%%%%%%%%%%%%%%%%%%%%%%%%%%%%%%%%%%%%%%%%%%%%%%
\subsection{Entanglement entropy for gauge theories on the world volumes of Dp branes }\label{higher}
%%%%%%%%%%%%%%%%%%%%%%%%%%%%%%%%%%%%%%%%%%%%%%%%%%%%%%%%%%%%%%%%%%%%%%%%%%%

\begin{figure}[!h]
		\subfigure[]{\label{fig:2-a}\includegraphics[scale=.70]{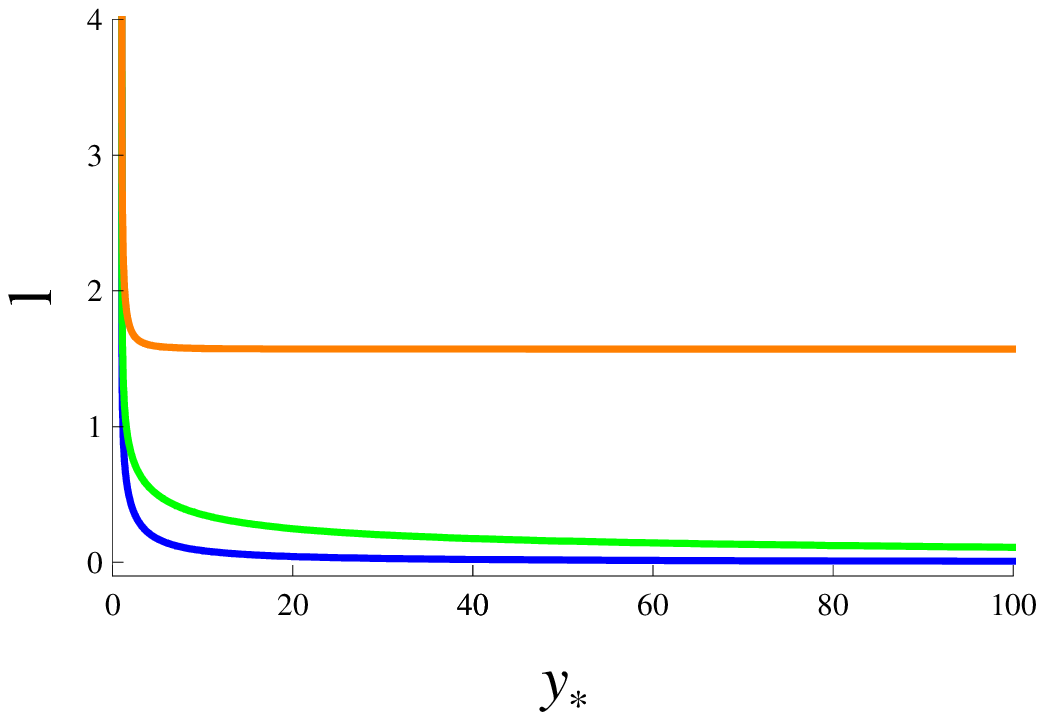}}
		\subfigure[]{\label{fig:2-b}\includegraphics[scale=.70]{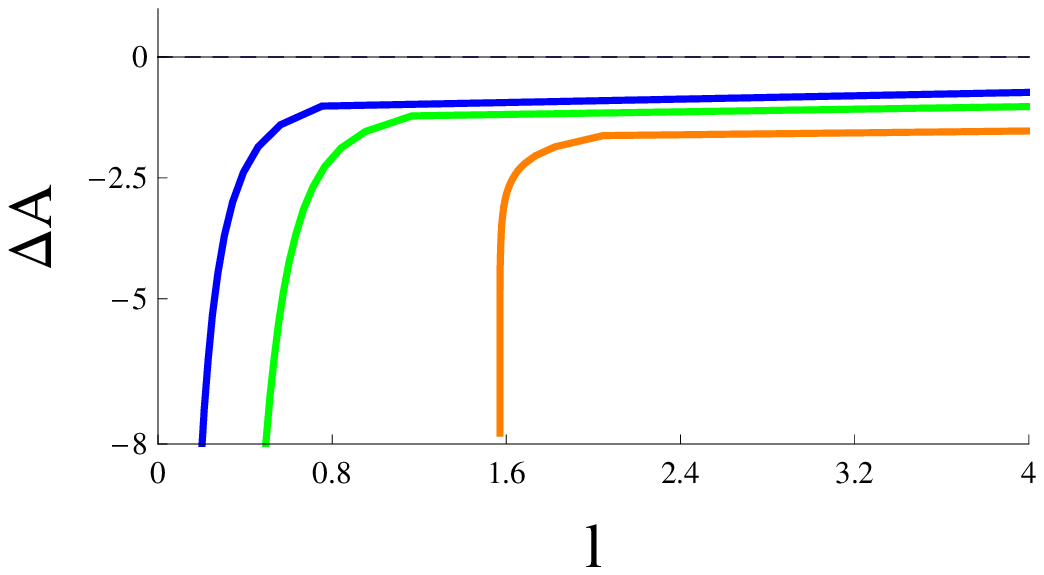}}
	\caption{Figure \ref{fig:2-a} shows the behavior of  $lu_0^{(p-5)/2}$ as a function of $y_*$ at $p=3,4,5$ (Blue, Green, Orange respectively).  Figure \ref{fig:2-b} Shows the difference in area as a function of $l$.  These plots show no phase transition for the entanglment entropy in the backgrounds of nonextremal D3, D4 and D5 branes.}
	\label{fig:finiteT}	
\end{figure}

\begin{figure}[!h]
		\subfigure[]{\label{fig:3-a}\includegraphics[scale=.70]{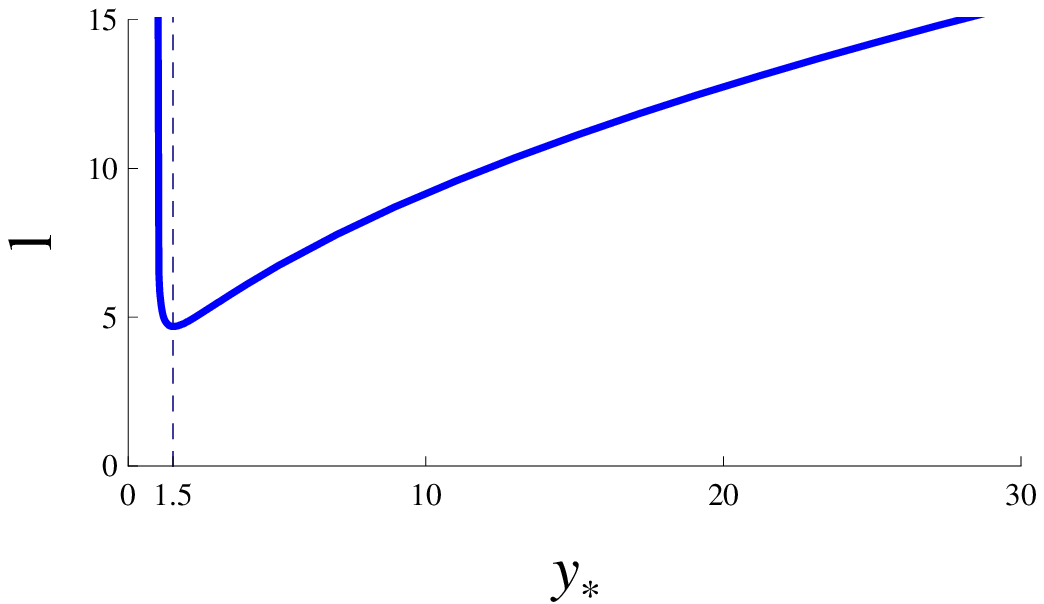}}
		\subfigure[]{\label{fig:3-b}\includegraphics[scale=.70]{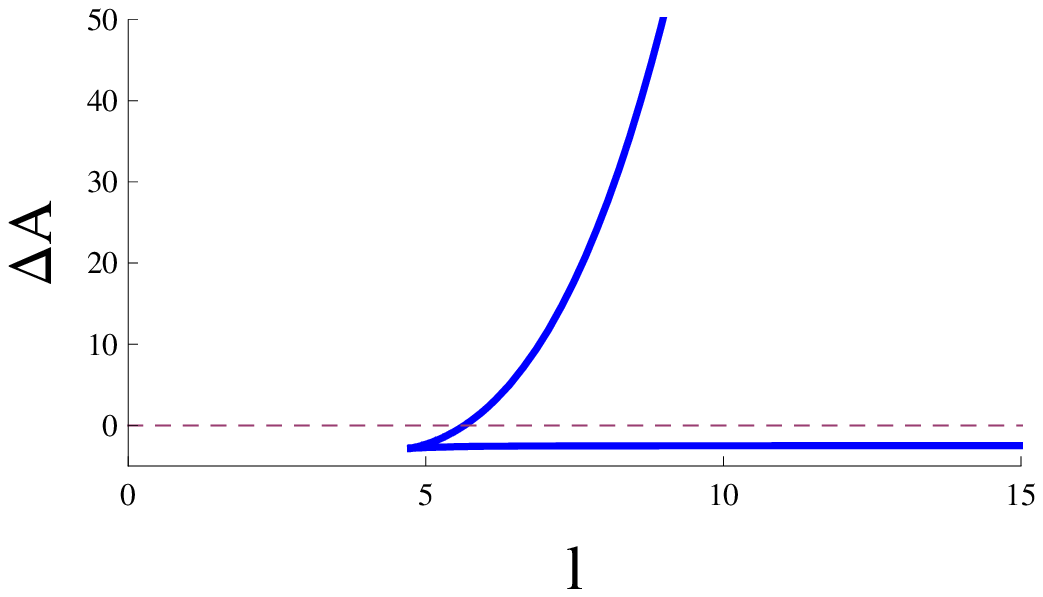}}
	\caption{Figure \ref{fig:finiteT6}(a) shows the behavior of  $l/\sqrt{u_0}$ as a function of $y_*$ for constant $u_0$ at $p=6$.  Figure \ref{fig:finiteT6}(b) Shows the difference in area as a function of $l$.  These plots show a transition for the entanglement entropy in the nonextremal D6 brackground.}
	\label{fig:finiteT6}	
\end{figure}

The backgrounds dual to finite temperature field theories are given in the string frame as \cite{Itzhaki:1998dd}:
\begin{equation}
    ds^2=\frac{u^{a/2}}{\sqrt{\lambda}}\Bigg[-\Bigg(1-\left(\frac{u_0}{u}\right)^a\Bigg)dt^2+dx_1^2+\cdots+dx_p^2\Bigg]+\frac{\sqrt{\lambda}}{u^{a/2}}\Bigg(1-\left(\frac{u_0}{u}\right)^a\Bigg)^{-1}du^2+\sqrt{\lambda}u^{b/2}d\Omega_{8-p}^2
\end{equation} where $a=7-p$ and $b=p-3$.  The dilaton is also given as:

\begin{equation}
    e^{\phi}=(2\pi)^{2-p}g_{YM}^2\left(\frac{\sqrt{\lambda}}{u^{a/2}}\right)^{-b/2}
\end{equation}  It is important to note that for $p<3$, the radius of the $(8-p)$-sphere diverges in the UV limit and shrinks in the IR limit.  However, for $p>3$, it shrinks to zero in the UV limit and expands in the IR limit.  At $p=3$, it decouples from the AdS sector.  A similar feature takes place for the dilaton.  The quantities $H$ and $\beta$ are
\begin{equation}
H(u) = \frac{C^2}{\lambda} u^{2\alpha}, \;\;\; \beta^2(u) = \frac{\lambda}{u^a - u_0^a}
\end{equation} where
\begin{equation}
C=\frac{(2\pi)^{2(p-2)}}{g_{YM}^4} \lambda \Omega_{8-p}, \;\;\; \alpha=\frac{1}{2}(9-p).
\end{equation}
The difference in areas $\Delta A$ and $l$ as integrals over $y$ are
\begin{eqnarray}
\Delta A = A_c - A_d &=& 2 C u_0^2 \left[ \int_{y_*}^{y_{\infty}}\frac{y^{2\alpha} -y^\alpha \sqrt{y^{2\alpha}-y_*^{2\alpha}} - y_*^\alpha}{\displaystyle{\sqrt{\left(y^{2\alpha}-y_*^{2\alpha}\right)\left(y^a-1\right)}}}dy -\int_{1}^{y_*}\frac{y^{\alpha}\,dy}{\sqrt{y^a-1}} \right] \nonumber \\
l&=& \frac{2\sqrt{\lambda}}{u_0^{\alpha-2}}\int_{y_*}^{y_{\infty}}\frac{y_*^{\alpha}dy}{\displaystyle{\sqrt{\left(y^{2\alpha}-y_*^{2\alpha}\right)\left(y^a-1\right)}}}.
\end{eqnarray}
We observe something interesting here. The difference in areas $\Delta A$ scales as $u_0^2$ for all $p$ while $l$ scales as $u_0^{\frac{p-5}{2}}$.  For $p>5$, it is proportional to $u_0$.  In figure \ref{fig:finiteT} we plot the difference in areas in units of $C u_0^2$ with respect to $l$ in units of $\sqrt{\lambda} u_0^{\frac{p-5}{2}}$ for $p = 3,4,5$.  In figure \ref{fig:finiteT6} we show the plot for $p=6$.

Here we observe a transition for $p=6$ and no transition for $p<6$.  This result is also in agreement with result \ref{alim}.

%%%%%%%%%%%%%%%%%%%%%%%%%%%%%%%%%%%%%%%%%%%%%%%%%%%%%%%%%%%%%%%%%%%%%%%%%%%
\section{Dyonic Black Hole} \label{dyonic}
%%%%%%%%%%%%%%%%%%%%%%%%%%%%%%%%%%%%%%%%%%%%%%%%%%%%%%%%%%%%%%%%%%%%%%%%%%%
This is a very important system from the holographic point of view. It has the potential to describe a few systems that are certainly of interest from the condensed matter point of view. A partial list of interesting applications can be found in \cite{Hartnoll:2007ih}.

\subsection{The solution }
The dyonic black hole is a solution to Einstein Maxwell on $AdS_4$.  The solution is a consistent truncation of 11 dimensional supergravity on $AdS_4\times S^7$.  Some interesting properties of this solution and its potential applications to condensed matter have been recently discussed in various papers including, \cite{Herzog:2007ij,Hartnoll:2007ai,Hartnoll:2007ih}

The relevant metric is:
\begin{equation} ds^2 = R^2 u^2 (-h(u) dt^2 + dz^2 + dy^2) + \frac{R^2}{u^2}\frac{du^2}{h(u)}, \end{equation}  with
\begin{equation} h(u) = 1 + (h^2 + q^2) \frac{u_0^4}{u^4} - (1 + h^2 + q^2) \frac{u_0^3}{u^3}, \end{equation} and electromagnetic field tensor \begin{equation} F = hu_0^2 dx\wedge dy - q u_0^2 dt\wedge du^{-1}. \end{equation} The three parameters $u_0$, $h$ and $q$  are related to the physical quantities, mass, electric charge and magnetic field of the black hole in the following way,
\begin{eqnarray} M= \alpha (1 + h^2 + q^2) u_0^3, \;\; C = q u_0^2, \;\; \mbox{and}, \;\; B = h u_0^2 \end{eqnarray} respectively.  The quantity $\alpha$ is a constant independent of $h,q$ and $u_0$.  It is useful to also define the parameters $\rho^2 = h^2 + q^2$ and $Q^2 = \rho^2 u_0^4$ to characterize the effect of the electromagnetic field.  In terms of the physical parameters, $u_0$ and $\rho^2$ are given by \begin{eqnarray} \rho^2 = \frac{Q^2}{u_0^4},\;\;\; u_0^4 + Q^2 - m u_0 = 0 \label{u:1} \end{eqnarray} where $m=M/\alpha$.  A computation of the mass using holographic renormalization can be found in \cite{Hartnoll:2007ai}.  Equation $\ref{u:1}$ has a positive real root for $u_0$ when
\begin{equation} \left(\frac{Q^2}{3}\right)^3 \leq \left(\frac{m}{4}\right)^4 \end{equation} which implies
\begin{equation} \left(\frac{\rho^2}{3}\right)^3 \leq \left(\frac{1+\rho^2}{4}\right)^4. \end{equation}  From these relationship, we can expect extremality at $\rho^2 =3$.  Furthermore, we observe that even though $Q$ is bounded by the mass, $\rho^2$ is allowed to take on any values.  Thus if there is extremality at $\rho^2=3$, we can expect different thermodynamical description of space when $\rho^2\leq 3$ from when $\rho^2 > 3$.  This is discussed in the next section.

%%%%%%%%%%%%%%%%%%%%%%%%%%%%%%%%%%%%%%%%%%%%%%
\subsection{Comments on thermodynamics}

With the coordinate change $y=u/u_0$, $h(u)$ can be rewritten
\begin{eqnarray} h(y) &=& \frac{1}{y^4} \left( y^4 + \rho^2 - (1+\rho^2)y \right) = \frac{1}{y^4} \left( y^4-y + \rho^2(1 - y) \right) \nonumber \\
&=& \frac{(y-1)}{y^4} \left( y^3 +y^2 +y -\rho^2 \right) \equiv \frac{1}{y^4}f(y)\nonumber \\ &=& \frac{(y-1)(y-\rho_0)}{y^4} \left( y^2 +(1+\rho_0)y + \frac{\rho^2}{\rho_0} \right) \end{eqnarray} where $\rho_0$ satisfies \begin{equation} \rho^3_0 + \rho^2_0 + \rho_0 - \rho^2 =0. \label{r:1} \end{equation}  We observe that $h(y)$ has at least one positive zero at $y=1$ corresponding to $u=u_0$; and at most two positive zeros with $y=\rho_0$ corresponding to $u = \rho_0 u_0$.  We can understand this second zero by studying equation (\ref{r:1}).  We can rewrite it as
\begin{equation} \rho^2(\rho_0) = \rho^3_0 + \rho^2_0 +\rho_0. \end{equation}  First we observe that $\rho_0$ vanishes when $\rho^2=0$ and $\rho_0=1$ when $\rho^2=3$.  In addition, the function $\rho^2(\rho_0)$ is monotonic since its derivative has no real zeros.  Thus $\rho^2$ has a one to one relationship with $\rho_0$.  This justifies the above statement that $h(u)$ has at most two real zeros.  Since $\rho^2$ is not bounded, $\rho_0$ is also unbounded.  The inverse of this function $\rho_0(\rho^2)$ is then:
\begin{equation} \rho_0 = \frac{b^2 - b -2}{3b} \;\;\; \mbox{where} \;\;\; b = \left[\frac{7+27\rho^2 + 3\sqrt{3}\sqrt{3+14\rho^2+27\rho^4}}{2}\right]^{1/3}. \end{equation}
The following picture emerges for the position of the horizon $\rho_* u_0$.
\begin{equation} \rho_* =  \begin{cases} &1 \;\;\; \mbox{when} \;\;\; 0\leq \rho^2 \leq 3 \\ &\rho_0  \;\;\; \mbox{when} \;\;\; 3\leq \rho^2 < \infty \end{cases}. \end{equation}  The temperature of the horizon is given by the surface gravity,
\begin{equation} T = \frac{\kappa}{2\pi} = \frac{1}{2\pi}\sqrt{\frac{-(\nabla^\alpha\xi^\beta)(\nabla_\alpha\xi_\beta)|_{u=\rho_* u_0}}{2}} \end{equation} where $\xi$ is a timelike killing vector.  With $\xi^\alpha = \left(\frac{\partial}{\partial t}\right)^\alpha$, the temperature is,
\begin{equation} T = \frac{u_0}{4\pi} \begin{cases} & 3 - \rho^2 \;\;\; \mbox{for} \;\;\; 0\leq \rho^2 \leq 3 \\ &\frac{2 \rho_0^4 +\rho_0 + \rho^2 (\rho_0 -2)}{\rho_0^3} \;\;\; \mbox{for} \;\;\ 3 \leq \rho^2 < \infty \end{cases}. \end{equation}  We observe extremality at $\rho^2 =3$ as expected.
\begin{figure}[!h]
  \centering
		\includegraphics[width=0.45\linewidth]{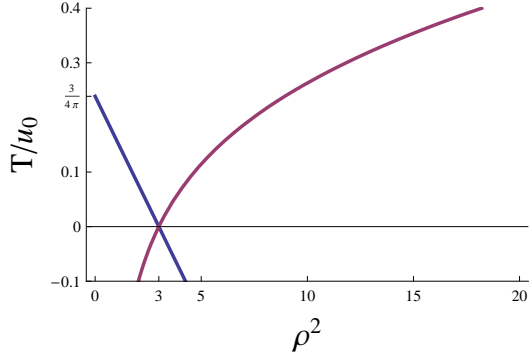}
	\caption{This plots shows the behavior of the temperature as a function of $\rho^2$.}
	\label{fig:dyotem}	
\end{figure}

In figure \ref{fig:dyotem}, the temperature is plotted against $\rho^2$.

\subsection{Entanglement Entropy}
The quantities $H$ and $\beta$ are:
\begin{equation}
H = R^2 u_0^2 y^2, \;\;\; \beta^2 = \frac{1}{u_0^2 f(y)}.
\end{equation} The difference in area and $l$ are given as
\begin{eqnarray}
\Delta A &=& 2 u_0 R^2 \left[ \int_{y_*}^{\infty}\frac{dy}{\sqrt{f(y)}}\left(\frac{y^4-\rho_*^2y_*^2}{\sqrt{y^4 - y^4_*}}-y^2\right) - \int_{\rho_*}^{y_*} \frac{y^2}{\sqrt{f(y)}} dy \right] \\
l &=& \frac{2 y_*^2}{u_0} \int_{y_*}^{\infty} \frac{1}{\sqrt{f(y)}}\frac{dy}{\sqrt{y^4 - y^4_*}}.
\end{eqnarray}
In the following graphs, $\Delta A$ is plotted in units of $u_0 R^2$ against $lu_0$ and $\rho^2$.

\begin{figure}[!h]
  \centering
		\includegraphics{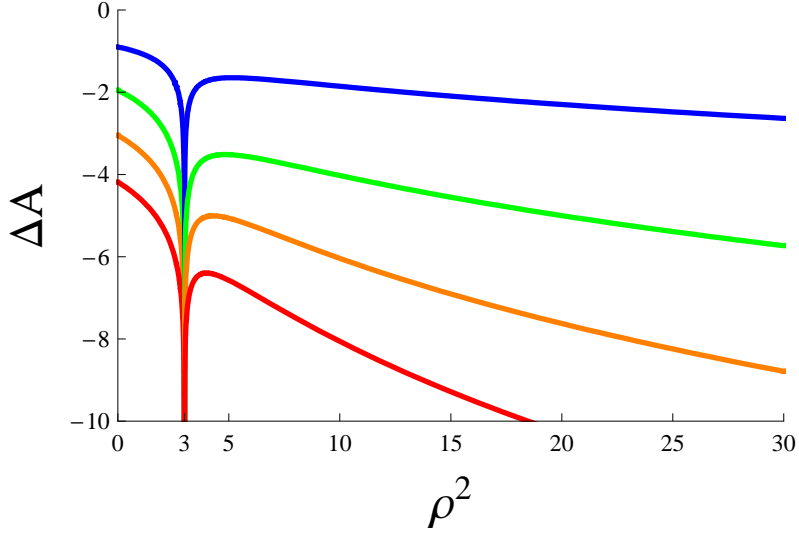}
	\caption{The difference in entropy is plotted against $\rho^2$ at $y_* = 1\rho_*,2\rho^2_*,3\rho_*,4\rho_*$ (Blue, Green, Orange and Red).  We observe that $\Delta A$ is always negative.}
	\label{fig:dyonicT}	
\end{figure}

\begin{figure}[!h]
		\subfigure[]{\label{fig:5-a}\includegraphics[scale=.70]{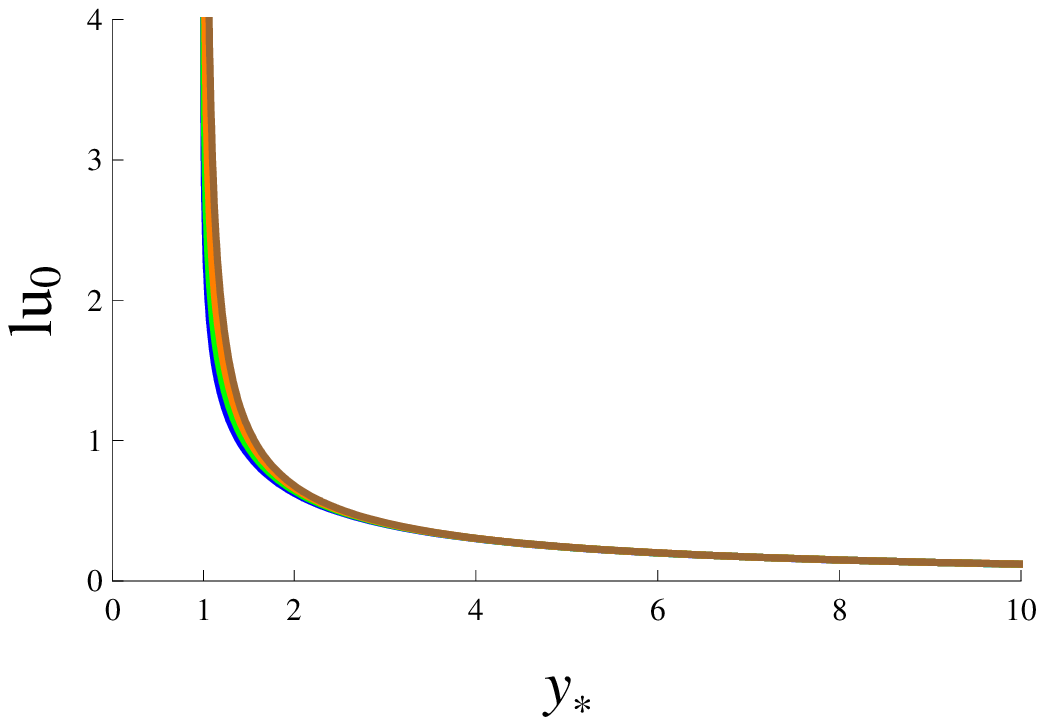}}
		\subfigure[]{\label{fig:5-b}\includegraphics[scale=.70]{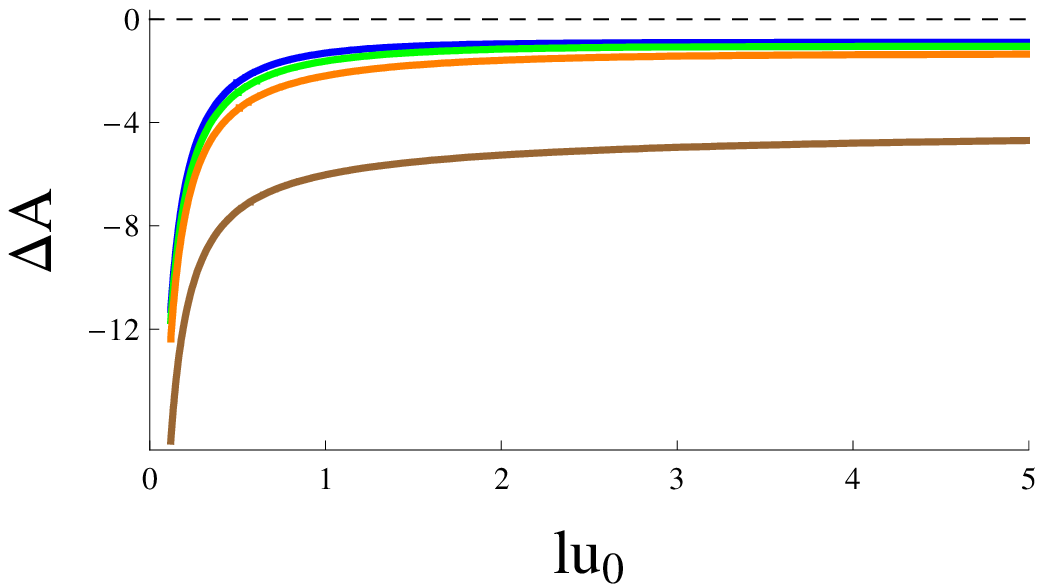}}
	\caption{Figure \ref{fig:dyonic*}(a) shows the behavior of the order parameter, $l$, as a function of $y_*$ for $\rho^2 = 0,1,2,2.99$ (Blue, Green, Orange, Brown).  Figure \ref{fig:dyonic*}(b) shows the difference in entropy as a function of the order parameter.  These plots show that there is no phase transition for $\rho^2<3$ in the dyonic black hole asymptotic to $AdS_4$.}
	\label{fig:dyonic*}	
\end{figure}

\begin{figure}[!h]
		\subfigure[]{\label{fig:6-a}\includegraphics[scale=.70]{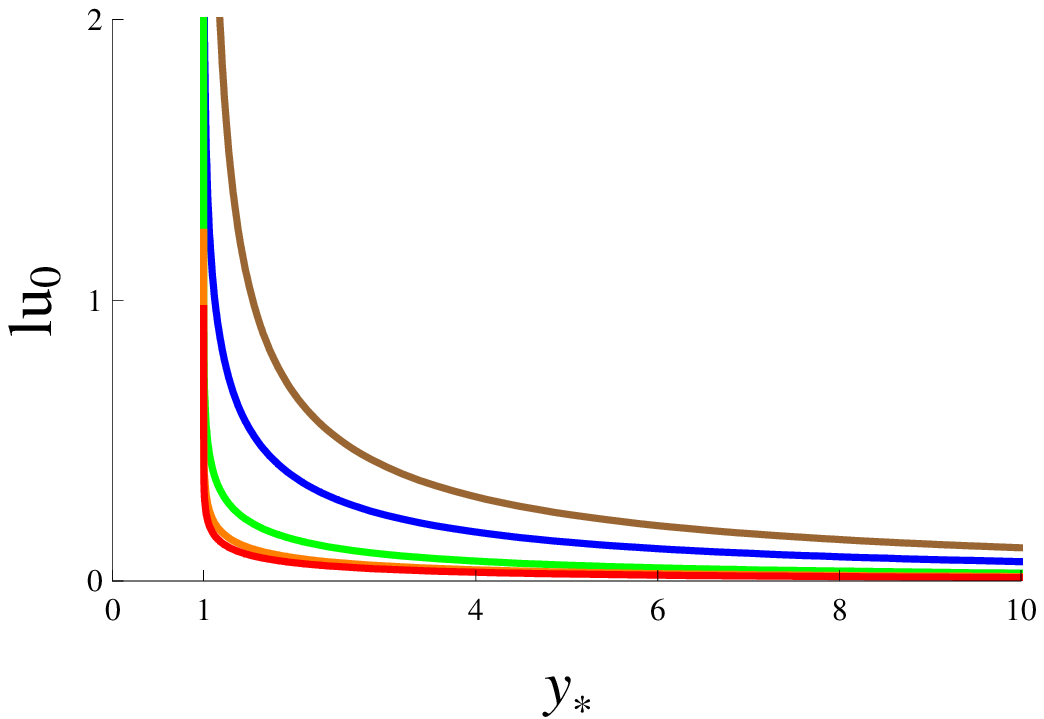}}
		\subfigure[]{\label{fig:6-b}\includegraphics[scale=.70]{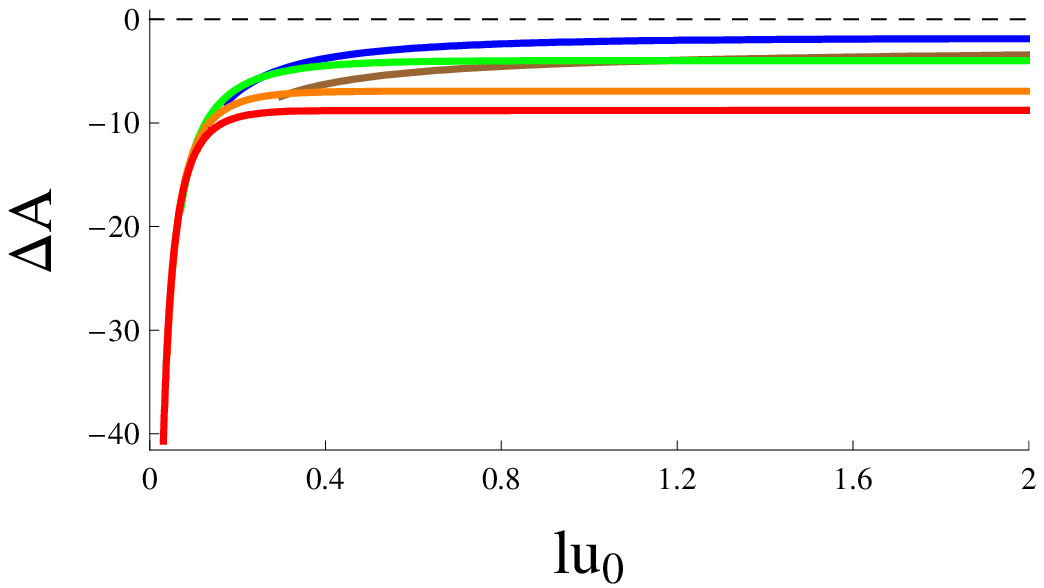}}
	\caption{Figure \ref{fig:dyonic*2}(a) shows the behavior of $l$ as a function of $y_*$ for $\rho^2 = 3.1,10,100,500,1000$ (Brown, Blue, Green, Orange, Red).  Figure \ref{fig:dyonic*2}(b) Shows the difference in entropy as a function of the order parameter. These plots show that there is no phase transition for $\rho^2>3$ in the dyonic black hole asymptotic to $AdS_4$.}
	\label{fig:dyonic*2}	
\end{figure}
In figure \ref{fig:dyonicT}, $\Delta A$ is plotted against $\rho^2$ for different values of $y_*$; here we observe that the difference in entropy is strictly bounded by zero from above.  In figure \ref{fig:dyonic*}(b), $\Delta A$ is plotted against $lu_0$ for values $\rho^2$ less than 3.  From this graph it is apparent that the difference in entropy approaches a constant negative value for large values $lu_0$.  We also note that an increase in $\rho^2$, translates to a decrease in $\Delta A$.  This feature is apparent in figure \ref{fig:dyonicT} and in figure \ref{fig:dyonic*}.  Furthermore, we also see that increasing $y_*$, in figure \ref{fig:dyonicT}, shifts $\Delta A$ downward. This allows us to also see that the difference in area is negative for large values of $l$; since small values of $y_*$ coincide with large values of $l$ as shown in figure \ref{fig:dyonic*}(a).  From these observations, we conclude that the difference in entropy is always negative for $\rho^2<3$.  Thus there is no transition.

For $\rho^2>3$, we also do not observe any transition.  This can be seen in figure \ref{fig:dyonicT} where the difference in entropy is strictly negative.  In figure \ref{fig:dyonic*2}(a) the difference in entropy is plotted against $lu_0$.  Here we observe an increase in entropy for small $\rho^2$ and then a monotonous decrease that is consistent with figure \ref{fig:dyonicT}.

%%%%%%%%%%%%%%%%%%%%%%%%%%%%%%%%%%%%%%%%%%%%%%%%%%%%%%%%%%%%%%%%%%%%
\section{Global $AdS_p$}\label{globalads}
%%%%%%%%%%%%%%%%%%%%%%%%%%%%%%%%%%%%%%%%%%%%%%%%%%%%%%%%%%%%%%%%%%

\subsection{Global $AdS_5$}
Now we explore the entanglement entropy of black holes in $AdS_5$ in global coordinates.  This geometry does not satisfy the conditions of section \ref{general} since there is no spacelike killing vector that commutes will with all killing vectors.  We reformulate the problem explicitly in this case. The Schwarzschild black hole in global $AdS_5$ is given by:
\begin{equation}
ds^2 = -f(r) dt^2 + \frac{1}{f(r)}dr^2 + r^2(d\theta^2 + \cos^2(\theta)d\phi^2 + \sin^2(\theta)d\varphi^2)
\end{equation}
where the 3-sphere is written in Hopf coordinates and $f(r)$ given by:
\begin{equation}
f(r) = 1-\frac{M}{r^2} + \frac{r^2}{L^2}.
\end{equation}
The conformal boundary is obtained, in these coordinates, by taking the $r\rightarrow \infty$ limit. The region $A$  is given by the hypersurface on the boundary parametrized as:
\begin{equation}
t = \mbox{constant} \;\;\; \mbox{and} \;\;\; 0 \leq \phi \leq 2\pi l.
\end{equation}  So, $l$ measures the size of $A$.  This quantity is bounded by 1 since the period of $\phi$ is $2\pi$.

\subsubsection{Entanglement}

The surfaces in $M$ that have the same boundary as $A$ can be parametrized as:
\begin{equation}
t=\mbox{constant} \;\;\; \mbox{and} \;\;\; r = r(\phi),
\end{equation}
with boundary conditions as:
\begin{equation}
r(0) = r(2\pi l) = r_\infty,
\end{equation} where $r_\infty$ is a cutoff parameter which will be taken to $\infty$.  The induced metric for this surface is given by:
\begin{equation}
ds^2 =  (\frac{r'^2(\phi)}{f(r)}+ r^2\cos^2(\theta))d\phi^2 + r^2(d\theta^2  + \sin^2(\theta)d\varphi^2).
\end{equation}
Its area is then
\begin{eqnarray}
A_c &=& 2\pi \int_0^{2\pi l}\int_0^{\pi/2} r^3 \sqrt{\cos^2(\theta) + b^2}\,\sin(\theta)\, d\theta \, d\phi \;\;\; \mbox{where} \;\;\; b=\frac{r'(\phi)}{r\sqrt{f(r)}} \\
&=& 2\pi \int_0^{2\pi l}\int_0^{1} r^3 \sqrt{a^2 + b^2} da  \, d\phi \;\; \mbox{where} \;\; a =\cos(\theta).
\end{eqnarray}
The minimal surface is obtained by solving for the function $r(\phi)$ that minimizes the action $A_c$.

$A_c$ is an action for a point particle with Lagrangian:
\begin{equation}
L = 2 \pi \int_0^{1} r^3 \sqrt{a^2 + b^2} \, da.
\end{equation}
From the Euler-Lagrange equation (H=Hamiltonian):
\begin{equation}
\frac{d}{d\phi} H = - \frac{\partial}{\partial \phi} L
\end{equation}
We obtain the equation of motion:
\begin{equation}
0 = \frac{d}{d\phi} \left[ (r'\frac{\partial}{\partial r'} -1)L \right] \, \Rightarrow \, (r'\frac{\partial}{\partial r'} -1)L = c
\end{equation} where $c$ is some constant.
Now we have
\begin{eqnarray}
r' \frac{\partial}{\partial r'}L -L &=& r^3 \int_0^1 \left[\frac{b^2}{\sqrt{a^2+b^2}}-\sqrt{a^2+b^2}\right] da \\
&=& - r^3 \int_0^1  \frac{a^2}{\sqrt{a^2+b^2}} da = -r^3 \int_{0}^{1} a \frac{d}{d a}\sqrt{a^2 + b^2} da \nonumber \\
&=& -r^3 \left[\sqrt{1+b^2} - \int_0^1\sqrt{a^2 + b^2} da \right]. \label{eq:a1}
\end{eqnarray} We can determine $c$ by considering the point $r_*$ where $r'=0$.  This corresponds to $b=0$. It is also important to note that $r_*$ must be bounded by $r_0$. We thus have:
\begin{equation}
c= -\frac{r_*^3}{2}.
\end{equation}  After integrating, we obtain the equation of motion of $r(\phi)$,
\begin{equation}
\frac{r_*^3}{r^3} = \sqrt{1+b^2} - b^2 \ln\left[\frac{1+\sqrt{1+b^2}}{b}\right].
\end{equation} We note that this is a transcendental equation for $b$.  Thus we cannot evaluate the area explicitly.  However we will be able to obtain some information when we substitute the equation of motion into the area formula.

Before we proceed, we first evaluate the piece-wise smooth surface.  This surface is the sum of 3 surfaces given by $\phi = 0$, $r= r_0$ and $\phi = 2\pi l$ at constant $t$.  The line element for the $\phi=$constant is,
\begin{equation}
ds^2 = \frac{1}{f(r)}dr^2 + r^2(d\theta^2 +\sin^2(\theta)d\varphi^2)
\end{equation}
with area
\begin{equation}
A = 2\pi \int_{r_0}^{\infty} \int_0^{\pi/2}  \frac{r^2}{\sqrt{f(r)}} \sin(\theta)\,d\theta dr = 2 \pi \int_{r_0}^{\infty} \frac{r^2}{\sqrt{f(r)}} dr.
\end{equation}

The $r=r_0$ piece has area
\begin{equation}
A = r_0^3 (2\pi)^2 l \int_0^{\pi/2} \sin{\theta}\cos{\theta} d\theta = 2\pi^2 r_0^3 l  .
\end{equation}
The area of the piece-wise smooth surface is then
\begin{equation}
A_d= 4 \pi \int_{r_0}^{\infty} \frac{r^2}{\sqrt{f(r)}} dr +  2\pi^2 r_0^3 l.
\end{equation}

\subsubsection{Comparison}

Before we can compare the areas, we rewrite the area of the smooth surface as an integral over $r$.  This is done by integrating out $a$; and then substituting $\frac{dr}{r'(\phi)}$ for $d\phi$ and integrated from $r_*$ to $r_\infty$.  One then obtains,

\begin{eqnarray}
A_c &=& \pi \int_{0}^{2 \pi l} r^3 \left[\sqrt{1+b^2} + b^2 \ln\left(\frac{1+\sqrt{1+b^2}}{b}\right)\right] d\phi \\
&=& 2 \pi \int_{r_*}^{r_\infty} \frac{r^2}{b\sqrt{f}} \left[\sqrt{1+b^2} + b^2 \ln\left(\frac{1+\sqrt{1+b^2}}{b}\right)\right] dr.
\end{eqnarray} We can also write $l$ as a function of $r_*$.  This is done by integrating $r'(\phi)$.
\begin{equation}
2 \pi l = \int_{0}^{2\pi l} d\phi = 2 \int_{r_*}^{r_\infty} \frac{dr}{r'} = 2 \int_{r_*}^{r_\infty} \frac{dr}{br\sqrt{f(r)}}.
\end{equation} Now we can evaluate the difference in area $\Delta A =A_c -A_d$.  It is given as:

\begin{eqnarray}
\Delta A &=& 2 \pi \int_{r_*}^{r_\infty} \frac{r^2}{b\sqrt{f}} \left[\sqrt{1+b^2} + b^2 \ln\left(\frac{1+\sqrt{1+b^2}}{b}\right)\right] dr \nonumber \\
&-& 4 \pi \int_{r_0}^{\infty} \frac{r^2}{\sqrt{f(r)}} dr - 2\pi r_0^3 \int_{r_*}^{r_\infty} \frac{dr}{br\sqrt{f(r)}} \label{eq:a1p}\\
&=& 2 \pi \int_{r_*}^{r_\infty} \frac{r^2}{b\sqrt{f}} \left[\sqrt{1+b^2} + b^2 \ln\left(\frac{1+\sqrt{1+b^2}}{b}\right) -\frac{r_0^3}{r^3} -2 b\right] dr \nonumber \\ &-& 4 \pi \int_{r_0}^{r_*} \frac{r^2}{\sqrt{f(r)}} dr.
\end{eqnarray}
Notice that we have split the second integral in equation \ref{eq:a1} to two integrals with ranges $r_0 \to r_*$ and $r_* \to r_\infty$.  Now we can use the equation of motion:
\begin{equation}
\frac{r_*^3}{r^3} = \sqrt{1+b^2} - b^2 \ln\left[\frac{1+\sqrt{1+b^2}}{b}\right]
\end{equation} to obtain
\begin{equation}
\Delta A = 2 \pi \int_{r_*}^{r_\infty} \frac{r^2}{b\sqrt{f}} \left[2b^2 \ln\left(\frac{1+\sqrt{1+b^2}}{b}\right) -2b   + \frac{r_*^3-r_0^3}{r^3}\right] dr - 4 \pi \int_{r_0}^{r_*} \frac{r^2}{\sqrt{f(r)}} dr.
\end{equation}
Now we can take the $r_*=r_0$ limit and obtain:
\begin{equation}
\Delta A = 4 \pi \int_{r_0}^{r_\infty} \frac{r^2}{\sqrt{f}} \left[b \ln\left(\frac{1+\sqrt{1+b^2}}{b}\right) -1\right] dr < 0
\end{equation} since
\begin{equation}
b \ln\left(\frac{1+\sqrt{1+b^2}}{b}\right) -1 < 0
\end{equation}
for nonzero values of $b$.  Thus we obtain
\begin{equation}
\Delta A(r_* \to r_0) < 0.
\end{equation}
We can also evaluate the large $r_*$ limit by eliminating the $\ln$ term in the difference equation:
\begin{equation}
\Delta A = 4 \pi \int_{r_*}^{r_\infty} \frac{r^2}{b\sqrt{f}} \left[\sqrt{1+b^2}- b -\frac{r_*^3 + r_0^3}{2r^3} \right] dr  - 4 \pi \int_{r_0}^{r_*} \frac{r^2}{\sqrt{f(r)}} dr. \label{dA:3}
\end{equation} As $r_* \rightarrow r_\infty$, we have
\begin{equation}
\Delta A = -4 \pi \int_{r_*}^{r_\infty} \frac{r_*^3}{2 r b\sqrt{f}} dr  - 4 \pi \int_{r_0}^{r_*} \frac{r^2}{\sqrt{f(r)}} dr.
\end{equation} Thus we obtain
\begin{equation}
\Delta A(r_* \rightarrow r_\infty) < 0.
\end{equation} Now we see that the difference in entropy is negative for both large values of $r_*$ and for $r_* = r_0$.  We collect all this information in:
\begin{equation}
\Delta A(r_*=r_0) < 0, \;\;\; \mbox{and} \;\;\; \Delta A(r_* \rightarrow r_\infty) < 0.
\end{equation}  From this argument, it is clear that if a transition exist, it must be at some intermediate value of $r_*$.  Furthermore, if there is transition, there exist a $r_*$ such that $\frac{d \Delta A (r_*)}{dr_*} = 0$.  This may not be straightforward to do since $\Delta A$ is an integral equation over $b(r,r_*)$; which satisfies a transcendental equation.  So one should integrate over $b$ instead of $r$ since the equation of motion can be read as:
\begin{equation}
r(b,r_*) = r_* \left( \sqrt{1+b^2} - b^2 \ln\left[\frac{1+\sqrt{1+b^2}}{b}\right] \right)^{-1/3}.
\end{equation}

One should also check that $r$ has a one to one relation with $b$, which can be done by showing that the first derivative has no zeros.  The derivative is then:
\begin{equation}
\frac{dr}{db} = r_* \frac{ 2b(-1-\sqrt{1+b^2} +(1+b^2 +\sqrt{1+b^2})\ln(\frac{1+\sqrt{1+b^2}}{b})) }{3\sqrt{1+b^2} (1+\sqrt{1+b^2})(\sqrt{1+b^2} -b^2 \ln(\frac{1+\sqrt{1+b^2}}{b}))^{4/3} }.
\end{equation}
Now, $db/dr$ is zero when the above expression diverges.  However, this expression is always finite making it safe to integrate over $b$ instead of $r$.  The integration bounds for $b$ will be from $0$ to $\infty$.  Notice that we have written $\Delta A$  in 3 different ways by using the equation of motion.  In what follows, we use the one given in equation (\ref{dA:3}).  It is
\begin{equation}
\Delta A = 4 \pi \int_{r_*}^{r_\infty} \frac{r^2}{b\sqrt{f}} \left[\sqrt{1+b^2}- b -\frac{r_*^3 + r_0^3}{2r^3} \right] dr  - 4 \pi \int_{r_0}^{r_*} \frac{r^2}{\sqrt{f(r)}} dr.
\end{equation}
It will also be useful to factor out the AdS radius $L$ by making the coordinate change $x=r/L$ and to introduce the parameter $m=M/L^2$.  We then have:
\begin{eqnarray}
f(r)&=& 1 - \frac{M}{r^2} + \frac{r^2}{L^2} = \frac{1}{x^2} (x^2 -x_0^2)(x^2 +x_0^2 +1) \equiv \frac{1}{x^2} h(x,x_0) \label{h:1}\\
 x_0^2 &=& \frac{\sqrt{1+4m} -1}{2} \\
x(b,y_*) &=& x_* \left( \sqrt{1+b^2} - b^2 \ln\left[\frac{1+\sqrt{1+b^2}}{b}\right] \right)^{-1/3} \\
\Delta A (x_*,x_0) &=& 4\pi L^3 \int_{x_*}^{x_\infty} \frac{x^3}{b\sqrt{h(x,x_0)}} \left[\sqrt{1+b^2}- b -\frac{x_*^3 + x_0^3}{2x^3} \right] dx \nonumber \\  &-& 4 \pi L^3 \int_{x_0}^{x_*} \frac{x^3}{\sqrt{h(x,x_0)}} dx.
\end{eqnarray} Notice that we have introduced the polynomial function $h(x,x_0)$ in (\ref{h:1}).  As an integral over $b$, the difference in area is given by:
\begin{eqnarray}
\frac{\Delta A}{4 \pi L^3} &=& \int_{0}^{\infty} \frac{x_*^4 g^3(b) g'(b)}{b\sqrt{h(x_* g(b),x_0)}}[\sqrt{1+b^2}-b] db - \int_{x_0}^{x_*} \frac{x^3}{\sqrt{h(x,x_0)}} dx \nonumber \\ &-& \int_{0}^{\infty} \frac{x_*(x_*^3 + x_0^3) g'(b)}{2b\sqrt{h(x_* g(b),x_0)}} db \;\;\; \mbox{where} \\
g(b) &=& \left( \sqrt{1+b^2} - b^2 \ln\left[\frac{1+\sqrt{1+b^2}}{b}\right] \right)^{-1/3}.
\end{eqnarray} We can also write $l$ as an integral over $b$:
\begin{equation}
l = \frac{x_*}{\pi} \int_{0}^{\infty} \frac{g'(b)}{b \sqrt{h(x_* g(b),x_0)}} db
\end{equation}

\begin{figure}[!h]
		\subfigure[]{\label{fig:global-a}\includegraphics[scale=.70]{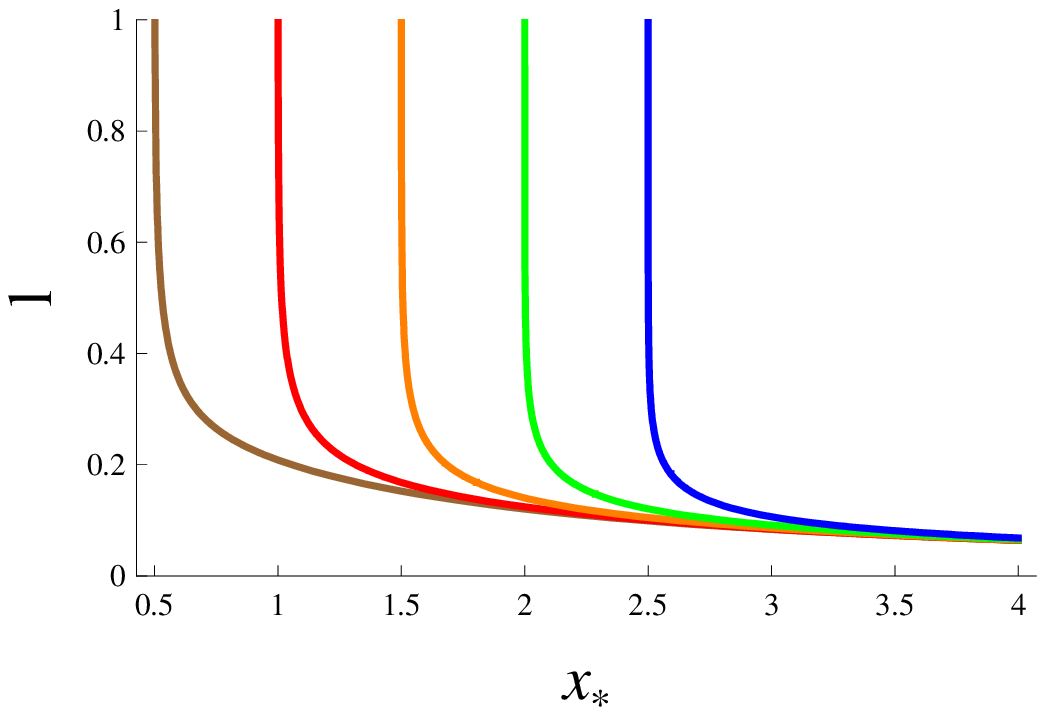}}
		\subfigure[]{\label{fig:global-b}\includegraphics[scale=.70]{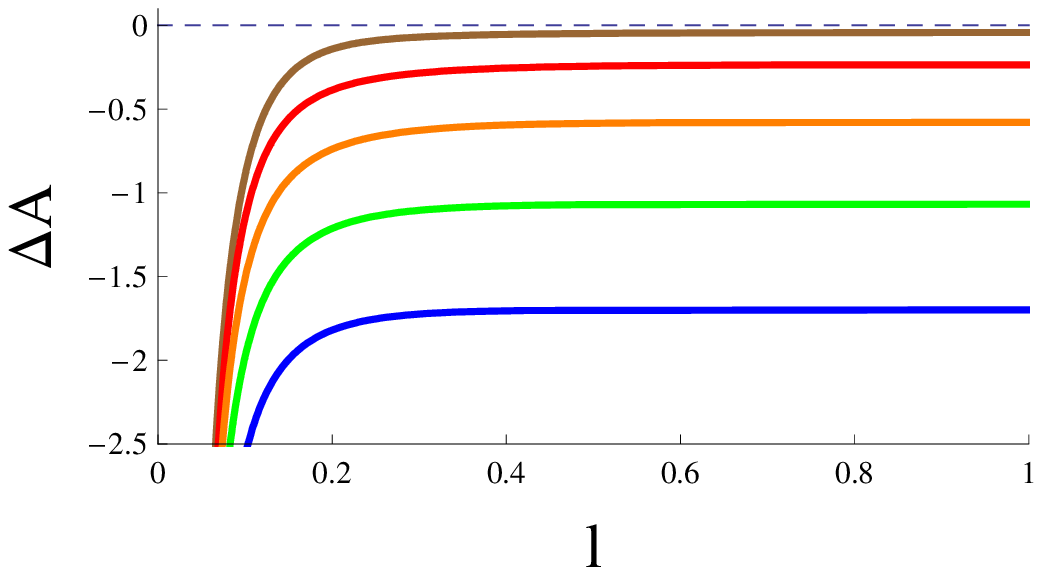}}
	\caption{Figure \ref{fig:global}(a) shows the behavior of the order parameter, $l$, as a function of $x_*$ for $x_0 = .5,1,1.5,2,2.5$ (Brown, Red, Orange, Blue).  Figure \ref{fig:global}(b) Shows the difference in entropy as a function of the order parameter.  These plots show no phase transition for the entanglement entropy in the Schwarzschild black hole in $AdS_5$.}
	\label{fig:global}	
\end{figure}

In figure \ref{fig:global} we have plotted the difference in area against $l$ at various horizon positions.  The features observed here are similar to those in the planar systems. However, we see that the difference in area shifts downward as we increase the temperature instead of just scaling.  Furthermore, we notice that there is a minimum $r_*$ for which one can define the entanglement entropy.  This minimum is given by the equation $l(r_*)=1$.

%%%%%%%%%%%%%%%%%%%%%%%%%%%%%%%%%%%%%%%%%%%%%%%%%%%%%%%%%%%%%%%%%%%%
\subsection{Global $AdS_4$} \label{global4}
%%%%%%%%%%%%%%%%%%%%%%%%%%%%%%%%%%%%%%%%%%%%%%%%%%%%%%%%%%%%%%%%%%

The relevant geometry is
\begin{equation}
ds^2 = -f(r) dt^2 + \frac{1}{f(r)}dr^2 + r^2(d\theta^2 + \sin^2(\theta)d\phi^2)
\end{equation}
where
\begin{equation}
f(r) = 1-\frac{M}{r} + \frac{r^2}{L^2}.
\end{equation}

Introducing the $x$-coordinate as  $x=r/L$, we find that the difference in area and the expression for $l$
\begin{eqnarray}
\Delta A &=& 2 L^2 \int_{x_*}^{x_\infty} \frac{dx}{b\sqrt{x}\sqrt{h(x,x_0)}} \left[x^2 (q(b)-\pi b) -2x_0^2\right] -2\pi L^2 \int_{x_0}^{x_*} \frac{dx x^{3/2}}{\sqrt{h(x,x_0)}} \\
l &=& \frac{1}{\pi} \int_{x_*}^{x_\infty} \frac{dx}{b\sqrt{x}\sqrt{h(x,x_0)}}.
\end{eqnarray} The equation of motion and $h(x,x_0)$ are given as
\begin{eqnarray}
x &=& x_* \frac{\sqrt{2}}{\sqrt{p(b)}} \;\;\; \mbox{where} \;\;\; b^2 = \frac{r'^2(\phi)}{r^2 f(r)}=\frac{x'^2(\phi)}{x^2 f(x,x_0)} \\
h(x,x_0) &=& x f(x,x_0) = (x-x_0)(x^2 + x x_0 +x_0^2 +1).
\end{eqnarray} The position of the horizon is given by $M$ as,
\begin{equation}
x_0^2 + x_0 -\frac{M}{L} = 0.
\end{equation} The functions $p(b)$ and $q(b)$ are:
\begin{eqnarray}
q(b) &=& \int_0^\pi \sqrt{b^2 + \sin^2(\theta)} d\theta \\
p(b) &=& \int_0^\pi \frac{\sin^2(\theta)}{\sqrt{b^2 + \sin^2(\theta)}} d\theta.
\end{eqnarray}
By substituting the equation of motion in the difference of area integral, one can see that
\begin{equation}
\Delta A (x_* \to x_0) < 0
\end{equation} since the integrand becomes negative for all $b$.  Now we can plot the above function by integrating over $b$.  In figure \ref{global4} we show the behavior of the difference in area with respect to $l$.  Again we do not observe a transition.

\begin{figure}[!h]
		\subfigure[]{\label{fig:global4-a}\includegraphics[scale=.70]{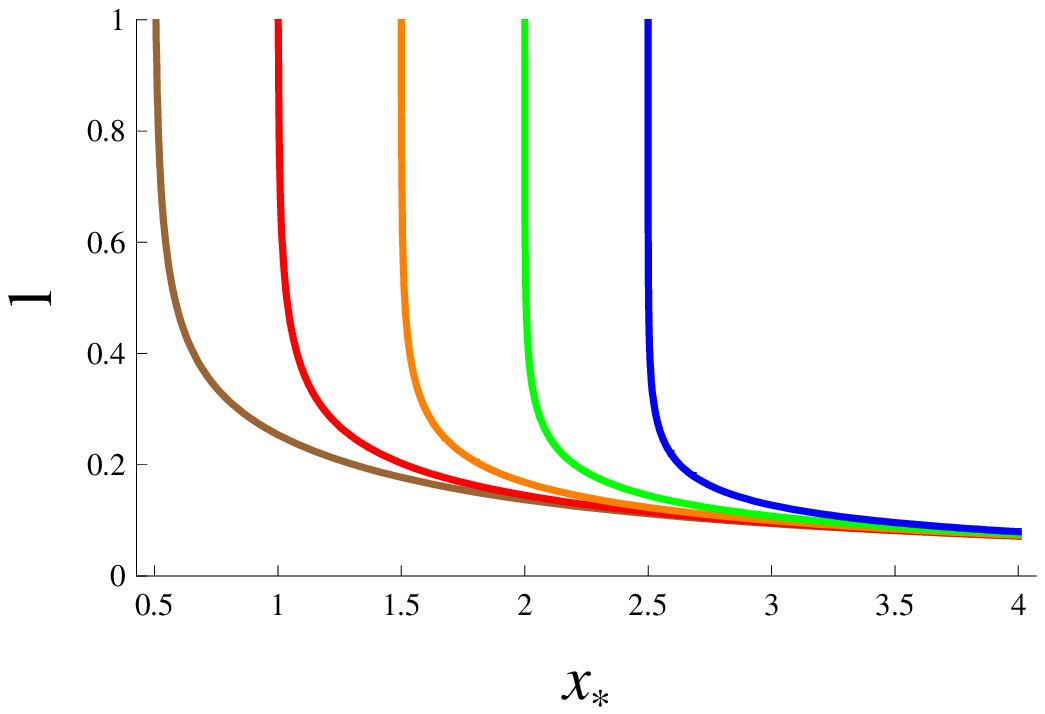}}
		\subfigure[]{\label{fig:global4-b}\includegraphics[scale=.70]{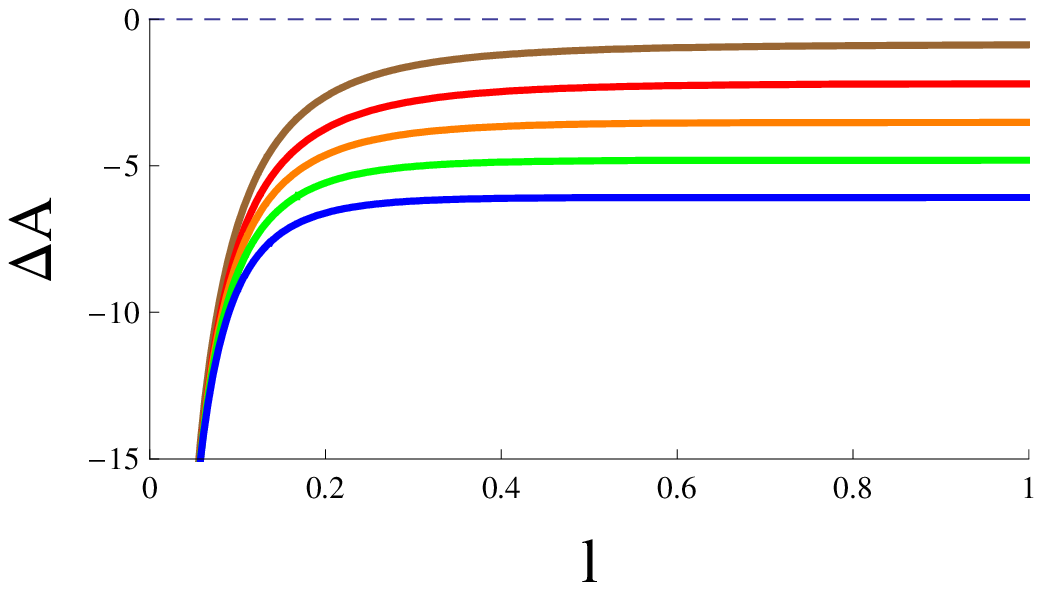}}
	\caption{Figure \ref{fig:global4}(a) shows the behavior of the order parameter, $l$, as a function of $x_*$ for $x_0 = .5,1,1.5,2,2.5$ (Brown, Red, Orange, Green, Blue).  Figure \ref{fig:global4}(b) Shows the difference in entropy as a function of the order parameter.  These plots show no phase transition for the entanglement entropy in the Schwarzschild black hole in $AdS_5$.}
	\label{fig:global4}	
\end{figure}

%%%%%%%%%%%%%%%%%%%%%%%%%%%%%%%%%%%%%%%%%%%%%%%%%%%%%%%%%%%%%%%%%%%%
\section{Conclusions}\label{conclusions}
%%%%%%%%%%%%%%%%%%%%%%%%%%%%%%%%%%%%%%%%%%%%%%%%%%%%%%%%%%%%%%%%%%

In lower dimensional systems the entanglement entropy has served as a way to detect exotic phases, see for example \cite{exotic}. In the context of higher dimensional field theories, most interestingly, four-dimensional field theories, it was recently suggested that the entanglement entropy could be used as an order parameter for confinement/deconfinement \cite{Klebanov:2007ws}. Motivated largely by these developments we launched a systematic study of the entanglement entropy in what should be understood as field theories at finite temperature. The hope was that by studying the entanglement entropy we would be able to identify a transition, just as a study of the entanglement entropy in the supergravity dual of the confined phase suggested the possibility of a transition to the deconfined phase. The premier examples of supergravity backgrounds dual to field theories at finite temperatures are the nonextremal Dp-brane solutions. We presented a detailed study in those geometries and generically detected no transition, except for the case of $p=6$ which is of limited interest from the field theory point of view.

We then analyzed the entanglement entropy in global coordinates. The motivation for the analysis comes from the fact that the Hawking-Page phase transition in asymptotically AdS spaces takes place only in global coordinates. This fact was given the interpretation, in the context of the AdS/CFT,  of being dual to the confinement/deconfinement transition for a field theory on a sphere. We did not observe a transition for most of the supergravity backgrounds. Since, as shown in \cite{Hawking:1982dh}, the free energy undergoes a transition in this context and the entanglement entropy does not we can safely conclude that the entanglement entropy is not the free energy. This is an important observation since the free energy is considered the standard order parameter for the confinement/deconfinement transition. The main reason arising from the natural expectation that in the confined phase the number of degrees of freedom comes from hadronic states and it is therefore of order $N^0$ for an $SU(N)$ theory while in the deconfined phase one expects $N^2$ degrees of freedom.

An interesting system where a Hawking-Page phase transition was established in \cite{Mahato:2007zm} is the black hole with cascading asymptotics constructed in \cite{PandoZayas:2006sa}. This system is the high temperature to which the Klebanov-Strassler background \cite{Klebanov:2000hb} flows. Preliminarily, with an analysis based on a few values of the temperature, we have been unable to observed a transition in the entanglement entropy for the background of \cite{PandoZayas:2006sa}. We differ a more detailed analysis to a separate publication.

The relationship between the entanglement entropy and the black hole entropy has been considered in various works in the literature, for example \cite{Brustein:2005vx,Emparan:2006ni,Kabat:1995eq,Kabat:1994vj}. In \cite{Kabat:1995eq} the one-loop corrections due to matter fields to the black hole entropy (computed through the free energy of the system) and to the entropy of entanglement were computed and found to agree for spin zero and spin one-half fields. A discussion more tuned to our main interest was presented in \cite{Brustein:2005vx}, where it was suggested that the microstates responsible for the black hole entropy are those due to the entanglement of the vacuum of the black hole. More bluntly, it was suggested in \cite{Brustein:2005vx} that the leading contribution to the black hole entropy is due to entanglement. A similar proposal, albeit with a major modification about the position of the black hole, was argued in \cite{Emparan:2006ni}.  The idea that black hole entropy is due to the entanglement entropy, in fact finds a place in our analysis and was also explicitly realized already in \cite{Ryu:2006ef} in the context of the holographic dual of the ${\cal N}=4$ plasma. We believe this is, ultimately, the right track but, as we have shown in various examples, the story is more complicated due to the fact that the entanglement entropy does not behave as an entropy obtained directly from the free energy. It is worth stressing the significant role of the horizon, in fact, if one does not include the contribution to the piece-wise smooth configuration coming directly from the horizon, there is generically a transition. So, we conclude that the horizon is crucial in the thermodynamical competition.

From a slightly more technical point of view, we can arrive at an important observation regarding collapsing cycles. In all these geometries, as in the cases discussed in  \cite{Klebanov:2007ws}, there is a  collapsing cycle coming from the temporal direction in Euclidean frame. However this cycle does not contribute in a significant way to the computations; in \cite{Klebanov:2007ws} the collapsing cycle was critical in obtaining transitions. An interesting alternative to the entanglement entropy has been recently suggested in \cite{Fujita:2008zv}. The geometric entropy defined in this paper uses a double Wick rotation and therefore includes the full effect of the collapsing Euclidean cycle that was, in the original geometry, the temporal direction. We predict that this modification will lead to transitions in most of the cases presented in this paper.

Finally, we end with a remark about the place of the entanglement entropy. We believe it is a thermodynamical potential rather than an entropy. Large parts of this intuition has been developed while collaborating with Carlos N\'u\~nez in similar issues. Namely, while studying the entanglement entropy, we observe that $l$  and $r_*$ (the minimum of the smooth surface) behave like thermodynamically conjugate variables.  This statement is motivated by the fact that when the function $l(r_*)$ is double valued, the entropy exhibits branches as observed in the geometries in \cite{Klebanov:2007ws} and for $p=6$.  This would suggest that the entropy is a new thermodynamical potential.  This is subject to further studies. Since we have observed, in a wide variety of models, no transition in the entanglement entropy we conclude that the entanglement entropy is, possibly,a thermodynamical potential that is different from the free energy. We are currently pursuing this line of investigation in collaboration with C. N\'u\~nez and expect to report on our findings soon.

%%%%%%%%%%%%%%%%%%%%%%%%%%%%%%%%%%%%%%%%%%%%%%%%%%%%%%%%%%%%%%%%%%%%
\section*{Acknowledgments}
%%%%%%%%%%%%%%%%%%%%%%%%%%%%%%%%%%%%%%%%%%%%%%%%%%%%%%%%%%%%%%%%%%
C. Terrero-Escalante is thankful to MCTP and  CEFIMAS for hospitality. This work is  partially supported by Department of Energy under
grant DE-FG02-95ER40899 to the University of Michigan.

%%%%%%%%%%%%%%%%%%%%%%%%%%%%%%%%%%%%%%%%

\end{document}